%% file: ml-paper.tex
\pdfoutput=1  % This tells arXiv to use pdflatex.

\documentclass[a4paper,11pt,twocolumn,twoside]{article}
\newlength{\MarginUnit}
\usepackage[vmargin=3\MarginUnit,hmargin=2\MarginUnit]{geometry}
\newcommand*{\Title}{Alternating Set Quantifiers in Modal Logic}
\newcommand*{\Author}{Fabian Reiter}
\newcommand*{\ShortAuthor}{F.~Reiter}
\newcommand*{\Email}{fabian.reiter@gmail.com}
\newcommand*{\Affiliation}{IRIF -- Universit\'e Paris Diderot}
\newcommand*{\Date}{February 2016}
\newcommand*{\Keywords}{Modal logic, Monadic second-order logic, Alternation hierarchies, Separation results, Graphs}

\usepackage{flushend}

\usepackage[fleqn]{amsmath}
\usepackage{amssymb}
\usepackage{mathtools}
\usepackage{stmaryrd}
\usepackage[nointegrals]{wasysym} % For the smiley :)

\usepackage{tikz}
\usepackage{standalone}
\usepackage{xcolor}

\usepackage{tabularx}  % for 'tabularx' environment and 'X' column type
\newcolumntype{Y}{>{\raggedright\arraybackslash}X}  % http://tex.stackexchange.com/a/194763
\usepackage{booktabs}  % for \toprule, \midrule, and \bottomrule macros
\newlength{\tablewidth}

\usepackage{caption}
\usepackage{enumitem}
\setlist{nosep,leftmargin=*}
\setlist[1]{topsep=1ex,itemsep=1ex}
\setlist[itemize,1]{label=\raisebox{0.25ex}{\tiny$\bullet$}}
\setlist[enumerate,1]{label=(\alph*)}

\usepackage[nottoc,notlot,notlof]{tocbibind}

\colorlet{darkblue}{blue!40!black}
\usepackage{hyperref}
\hypersetup{pdftitle={\Title}, 
            pdfauthor={\Author},
            pdfkeywords={\Keywords},
            colorlinks=true,
            allcolors=darkblue,
            linktoc=section} 
\usepackage[capitalise,noabbrev]{cleveref} % Must be loaded after hyperref.

\title{\huge{\Title}}
\author{
  \Author \\
  \small{\Affiliation} \\
  \href{mailto:\Email}{\footnotesize{\nolinkurl{\Email}}}
}
\date{\small{\Date}}

\renewenvironment{abstract}
{\vspace{-6ex}
  \setlength{\leftmargini}{3\MarginUnit}
  \quotation
  \noindent\rule{\linewidth}{\heavyrulewidth}
  \smallbreak
  {\bfseries\noindent{\abstractname.}\:}}
{\par
  \noindent\rule{\linewidth}{\heavyrulewidth}
  \endquotation
  \vspace{5ex}}

\usepackage{titleps}
\newpagestyle{mystyle}{
  \sethead[\textit{\ShortAuthor}][][]{}{\textit{\Title}}{}
  \setfoot{}{\thepage}{}
}
\usepackage{sty/typoconfig} % Must be loaded before sty/inputconfig.
\usepackage{sty/inputconfig}
\usepackage{sty/theoremconfig}
\usepackage{sty/commands}
\usepackage{sty/modallogic}

\begin{document}

\twocolumn[
  \begin{@twocolumnfalse}
    \maketitle
    \input{tex/abstract.tex}
  \end{@twocolumnfalse}
]

{\small\tableofcontents}

\input{tex/introduction.tex}
\input{tex/notation.tex}
\input{tex/results.tex}
\input{tex/proofs.tex}
\input{tex/grids.tex}
\input{tex/encodings.tex}
\input{tex/acknowledgments.tex}

\input{tex/references.tex}
\end{document}

%% file: tex/abstract.tex
\begin{abstract}
  We establish the strictness of several set quantifier alternation hierarchies
  that are based on modal logic,
  evaluated on various classes of finite graphs.
  This extends to the modal setting a celebrated result of Matz, Schweikardt and Thomas~(2002),
  which states that the analogous hierarchy of monadic second-order logic is~strict.
  
  Thereby, the present paper settles a question raised by van~Benthem~(1983),
  revived by ten~Cate~(2006),
  and partially answered by Kuusisto~(2008, 2015).
\end{abstract}

%% file: tex/introduction.tex
\section{Introduction}
\label{sec:introduction}
One of the central concerns in theoretical computer science
is to understand which components of any given formal system
are responsible for its expressive power.
A typical sort of question is thus
whether the power of a particular model decreases
when it is deprived of a certain feature.
In the present article,
we investigate two questions of that very sort.
As they have emerged independently
in separate areas of research, with different perspectives,
their relationship may not be obvious at first sight.

\subsection{Motivation and Related Work}
Our first question is an old problem in the field of modal logic.
Formulas in that system are usually evaluated on labeled graphs
from the point of view of a distinguished vertex,
which can only see the labels of its neighbors within a small radius.
A natural generalization of this model,
investigated in various places of the literature,
such as \cite{Bul69} and \cite{Fin70},
is to introduce quantifiers over sets,
by means of which a formula can, so to speak,
extend the vertex labeling of a given graph.
We shall refer to a variant of the resulting formalism as
\mbox{\emph{hybrid logic with set quantifiers} ($\HS$)}.
Already in 1983,
van~Benthem asked in \cite{Ben83}
whether the syntactic hierarchy
obtained by alternating between existential and universal set quantifiers
induces a corresponding hierarchy on the semantic side.
Remaining unanswered,
the question was raised again by ten~Cate in \cite{Cat06},
and finally a partial answer was provided
by Kuusisto in \mbox{\cite{Kuu08,Kuu15}}:
He could show that $\HS$
induces an \emph{infinite} hierarchy over finite directed graphs.
This tells us that the hierarchy does not completely collapse at some level,
but leaves open whether or not each number of quantifier alternations
corresponds to a separate semantic level.
Kuusisto's proof builds upon the work
of Matz, Schweikardt and Thomas in \cite{MST02}
(elaborating on their previous results in \cite{MT97} and \cite{Sch97}),
where they have shown that,
in the case of \emph{monadic second-order} ($\MSO$) logic,
the analogous hierarchy is \emph{strict}.
Thus,
each additional alternation between the two types of set quantifiers
properly extends the according family of definable graph properties.
Significantly,
this result also holds for a more restrictive class of structures called grids.

The second problem,
and the original motivation for the present article,
stems from the field of automata theory on graphs.
There, the author has introduced in \cite{Rei15} a notion of graph automaton,
dubbed \emph{alternating distributed graph automaton} (ADGA),
which operates in a manner similar to a distributed algorithm.
Equipped with the ability to alternate between
nondeterministic decisions of the individual processors
and the creation of parallel computation branches,
this model has been shown equivalent to $\MSO$ on graphs.
As a natural follow-up question,
one can ask whether successive restrictions on
the capability of alternating between the two operation modes
always lead to a decrease in expressive power.
This seems particularly relevant in light of
the corresponding separation results on $\MSO$
obtained by Matz, Schweikardt and Thomas.

As it turns out,
the two problems above are strongly related.
The reason is that
ADGA can be viewed as the automata-theoretic counterpart of $\HGS$,
a logic extending $\HS$ with an additional operator,
called \emph{global modality},
that allows quantification over all evaluation points.
Two corresponding levels of alternation in the frameworks of ADGA and $\HGS$
characterize exactly the same graph properties.
Although this observation has so far not been formally published,
it is relatively easy to verify, once pointed out.
Besides,
a similar correspondence between local distributed algorithms and modal logic
has already been established by Hella et~al.\ in \cite{H+12,H+15},
and the equivalence of $\HGS$ and $\MSO$
that can be indirectly inferred from \cite{Rei15}
has previously been shown in \cite{Kuu08,Kuu15}.

Hereafter,
we shall work with $\HGS$ instead of the ADGA model,
because compared to state diagrams,
logical formulas take up less space and are usually easier to manipulate.

\subsection{Contribution}
The present work gives a complete answer to both of the questions mentioned above
and to some variants thereof.
In particular,
the set quantifier alternation hierarchies induced by $\HS$ and $\HGS$
are shown to be \emph{strict} over finite directed graphs.
Just as Kuusisto has done in \cite{Kuu08,Kuu15},
we will use as a starting point
the strictness result of \cite{MST02} for $\MSO$ on grids.
But from there on,
the two proof methods diverge considerably.

Kuusisto's approach is mainly based on the fact
that one can simulate first-order quantifiers
by means of set quantifiers,
combined with a formula stating that a set is a singleton.
For $\HGS$,
this results in the mentioned equivalence with $\MSO$,
which immediately implies
that the hierarchy of $\HGS$ must be infinitely ascending.
The spirit of his proof remains the same for $\HS$,
although the details are much more technical.
It is precisely this use of additional second-order quantifiers
that leads to the loss of the specific separation results provided by~\cite{MST02}.

In contrast,
one crucial insight
will enable us to take full advantage of those results:
When restricted to the class of grids,
$\HGS$ and $\MSO$ are more than just equivalent~---
they are \emph{levelwise} equivalent,
and consequently
all the separation results shown for $\MSO$
also hold for $\HGS$ on grids.
This is based on the observation that the existential fragment of $\HGS$
can simulate another model, called \emph{tiling systems},
which has been shown to be equivalent to the existential fragment of $\MSO$
in \cite{GRST96}.
On the basis of this new finding,
we can then transfer the given separation results from $\HGS$ on grids
to other classes of graphs and other extensions of modal logic,
such as $\HS$.
While this works along the same general principle as the
\emph{strong first-order reductions} used in \cite{MST02},
the additional limitations imposed by modal logic
force us to introduce custom encoding techniques
that cope with the lack of resources.

\subsection{Outline}
The remainder of this paper is organized in a top-down manner.
After introducing the necessary terminology in \cref{sec:notation},
we present the main results in \cref{sec:results},
and almost immediately get to the central proof in \cref{sec:proofs}.
The latter relies on several other propositions,
but since those are treated as “black boxes”,
the main line of reasoning should be comprehensible
without reading any further.
We then provide all the missing details
in the last two \lcnamecrefs{sec:grids},
which are independent of each other.
\Cref{sec:grids} establishes the levelwise equivalence
of three different alternation hierarchies on grids,
and may thus be interesting on its own.
On the other hand,
\cref{sec:encodings} is dedicated to encoding functions,
which constitute the more technical part of our demonstration.

%% file: tex/notation.tex
\section{Notation and Terminology}
\label{sec:notation}
We begin by defining the basic vocabulary used throughout this paper.
It is mostly standard, with one noteworthy exception:
We shall not make the usual distinction between variables and (non-logical) constants.
Instead,
there is simply a fixed supply of symbols,
which can serve both as variables and as constants.
The set \defd{$\ElemSyms$} contains our \defd{element symbols},
among which there is a special position symbol~\defd{$\at$}.
Furthermore, for every positive integer $k$,
we let \defd{$\RelSyms[k]$} be the set of $k$-ary \defd{relation symbols}.
All of these sets are infinite and pairwise disjoint.
We also denote the set of all \defd{symbols} by $\Symbols$,
i.e., $\defd{\Symbols} \defeq \bigcup_{k≥0}\Symbols[k]$,
and shall often refer to the unary relation symbols in $\SetSyms$ as \defd{set symbols}.

\subsection{Structures}
Let $\Signature$ be any subset of $\Symbols$.
A \defd{structure} $\A$ of signature $\Signature$ consists of
a nonempty set \defd{$\domain{\A}$}, called the \defd{domain} of $\A$,\,
an element \defd{$\inp{\A}{q}$} of $\domain{\A}$ for each element symbol $q$ in~$\Signature$, and
a $k$-ary relation \defd{$\inp{\A}{R}$} on $\domain{\A}$ for each $k$-ary relation symbol $R$ in $\Signature$.
Here, $\inp{\A}{q}$ and $\inp{\A}{R}$ are called
$\A$'s \defd{interpretations} of the symbols $q$ and $R$.
We may also say that $\A$ is a structure \defd{over}~$\Signature$,
or that $\Signature$ is the \defd{underlying} signature of $\A$,
and we denote $\Signature$ by \defd{$\signature{\A}$}.
In case the position symbol~$\at$ lies in $\signature{\A}$,
we call $\A$ a \defd{pointed structure}.

As is customary,
we are only interested in structures up to isomorphism.
That is,
two structures over $\Signature$ are considered to be equal
if there is a bijection between their domains
that preserves the interpretations of all symbols~in~$\Signature$.

For convenience, we will often neglect the notational distinction
between a structure and its domain.
Hence, when we write \defd{$a∈\A$} and \defd{$A⊆\A^k$},
we mean $a∈\domain{\A}$ and $A⊆(\domain{\A})^k$, respectively.

We use the notation \defd{$\ver{\A}{S}{α}$}
to designate the structure $\A'$ obtained from $\A$ by interpreting the symbol $S$ as $α$,
where $α∈\A$ if $S∈\ElemSyms$, and $α⊆\A^k$ if $S∈\RelSyms[k]$.
More precisely, $\domain{\A'} = \domain{\A}$,\,
$\signature{\A'} = \signature{\A}∪\set{S}$,\,
$\inp{\A'}{S} = α$,\,
and $\inp{\A'}{T} = \inp{\A}{T}$ for $T∈\signature{\A}\setminus\set{S}$.
This is generalized to multiple symbols in the obvious way.
If the interpretation of $S$
is clear from context,
we may refer to the structure above as the \defd{$S$-extended variant of $\A$}.

\subsection{Different Kinds of Graphs}
Our main focus will be on several types of structures
with finite domains and relations of arity at most $2$.
In the following definitions, let $t$ and $u$ be non-negative integers.

A \defd{$t$-bit labeled, $u$-relational directed graph} $\sD$
is a finite structure of signature $\set{P_1,…,P_t,R_1,…,R_u}$,
where $P_1,…,P_t$ are set symbols,
and $R_1,…,R_u$ are binary relation symbols.
The class of all such structures is denoted by \defd{$\DIGRAPH[t,u]$}.
We sometimes refer to $\inp{\sD}{P_1},…,\inp{\sD}{P_t}$ as \defd{labeling sets}
and to $\inp{\sD}{R_1},…,\inp{\sD}{R_u}$ as \defd{edge relations}.
If the latter are all irreflexive and symmetric,
then $\sD$ is called a $t$-bit labeled, $u$-relational \defd{undirected graph},
and the corresponding class is \defd{$\GRAPH[t,u]$}.
Furthermore,
if $t=0$ and $u=1$, we say that $\sD$ is simply an \defd{(un)directed graph},
and use the shorthands
$\defd{\DIGRAPH} \defeq \DIGRAPH[0,1]$ and $\defd{\GRAPH} \defeq \GRAPH[0,1]$.
We shall also drop the subscripts, and just write $P$ or $R$,
if there is only one symbol of a given arity.

As can be easily guessed from the previous definitions,
a \defd{pointed directed graph} is a directed graph
in which some element has been marked by the position symbol~$\at$,
i.e., a structure of the form $\ver{\sD}{\at}{d}$,
with $\sD∈\DIGRAPH$ and $d∈\sD$.
We write \defd{$\PDIGRAPH$} for the set of all pointed directed graphs.

Finally,
we also consider an important subclass of $\DIGRAPH[t,2]$,
whose members represent rectangular labeled grids (also called pictures).
In such a structure~$\sC$,
each element is identified with a grid cell,
and the edge relations $\inp{\sC}{R_1}$ and $\inp{\sC}{R_2}$
are interpreted as
the “vertical” and “horizontal” successor relations, respectively.
The unique element that has no predecessor at all is regarded as the
“upper-left corner”,
and all the usual terminology of matrices applies.
Formally,
$\sC$ is a \defd{$t$-bit labeled grid} if, for some $m,n≥1$,
it is isomorphic to a structure
with domain $\set{1,…,m}×\set{1,…,n}$
and edge relations
\begin{align*}
  \inp{\sC}{R_1} &= \bigsetbuilder{\bigtuple{\tuple{i,j},\tuple{i+1,j}}}{1≤i<m,\,1≤j≤n}, \\[0.5ex]
  \inp{\sC}{R_2} &= \bigsetbuilder{\bigtuple{\tuple{i,j},\tuple{i,j+1}}}{1≤i≤m,\,1≤j<n}.
\end{align*}
If $t=0$, we refer to $\sC$ simply as a \defd{grid}.
In alignment with the previous nomenclature,
we let \defd{$\GRID$} and \defd{$\GRID[t]$} denote
the classes of grids and $t$-bit labeled grids.

\subsection{The Considered Logics}
As we shall contemplate both classical logic and numerous variants of modal logic,
we introduce them all in a common framework.
First we define the syntax and semantics of a generalized language,
and then we specify which particular syntactic fragments we are interested~in.

\begin{table*}[htbp]
  \centering
  \begin{tabularx}{\tablewidth}{llY}
    \toprule
    \textit{Syntax} & \textit{Free symbols} & \textit{Semantics} \\
    Formula $ψ$ & Symbol set $\free{ψ}$ & Necessary and sufficient condition for $\A⊨ψ$ \\
    \midrule\addlinespace
    $q$ & $\set{\at,q}$ & $\inp{\A}{\at} = \inp{\A}{q}$ \\\addlinespace
    $(p\eqsymb q)$ & $\set{p,q}$ & $\swl{\inp{\A}{p}}{\inp{\A}{\at}} = \inp{\A}{q}$ \\\addlinespace
    $P$ & $\set{\at,P}$ & $\inp{\A}{\at} ∈ \inp{\A}{P}$ \\\addlinespace
    $P(q)$ & $\set{q,P}$ & $\swl{\inp{\A}{q}}{\inp{\A}{\at}} ∈ \inp{\A}{P}$ \\\addlinespace
    $R(q_0,…,q_k)$ & $\set{q_0,…,q_k,R}$ & $\tuple{\inp{\A}{q_0},…,\inp{\A}{q_k}} ∈ \inp{\A}{R}$ \\\addlinespace
    $¬φ$ & $\free{φ}$ & not\, $\A⊨φ$ \\\addlinespace
    $(φ_1∨φ_2)$ & $\free{φ_1}∪\free{φ_2}$ & $\A⊨φ_1$ \,or\, $\A⊨φ_2$ \\\addlinespace
    $\dm[R](φ_1,…,φ_k)$ & $\set{\at,R}∪\,\smashoperator{\bigcup_{1≤i≤k}}\free{φ_i}$
                        & For some $a_1,…,a_k\mathbin{∈}\A$ such that $\tuple{\inp{\A}{\at},a_1,…,a_k}∈\inp{\A}{R}$, \newline
                          we have $\ver{\A}{\at}{a_i}⊨φ_i$ \mbox{for each $i∈\set{1,…,k}$}. \\\addlinespace[0.8\defaultaddspace]
    $\bdm[R](φ_1,…,φ_k)$ & same as above
                         & As above, except for the condition $\tuple{a_k,…,a_1,\inp{\A}{\at}}∈\inp{\A}{R}$. \\\addlinespace
    $\gdm φ$ & $\free{φ}\setminus\set{\at}$
             & $\swl{\ver{\A}{\at}{a}}{\ver{\A}{P}{A}}⊨φ$ for some $a∈\A$ \\\addlinespace
    $\EE{q}φ$ & $\free{φ}\setminus\set{q}$ 
              & $\swl{\ver{\A}{q}{a}}{\ver{\A}{P}{A}}⊨φ$ for some $a∈\A$ \\\addlinespace
    $\EE{P}φ$ & $\free{φ}\setminus\set{P}$ 
              & $\ver{\A}{P}{A}⊨φ$ for some $A⊆\A$ \\
    \addlinespace\midrule
    \multicolumn{3}{l}{Here,\, $p,q,q_0,…,q_k∈\ElemSyms$,\: $P∈\SetSyms$,\, $R∈\RelSyms[k+1]$,
                       and $φ,φ_1,…,φ_k$ are formulas, for $k≥1$.} \\
    \bottomrule
  \end{tabularx}
  \caption{Syntax and semantics of the considered logics.}
  \label{tab:syntax-semantics}
\end{table*}

\begin{table*}[htbp]
  \centering
  \newcommand*{\seplinespace}{\addlinespace[1.5\defaultaddspace]}
  \begin{tabularx}{\tablewidth}{lYl}
    \toprule
    \multicolumn{2}{l}{\textit{Language}} & \textit{Generating grammar} \\
    \midrule\addlinespace
    \defd{$\H$}   & basic Hybrid 
                  & $φ \Coloneqq q \mid P \mid ¬φ \mid (φ_1∨φ_2) \mid \dm[R](φ_1,…,φ_k)$ \\\seplinespace
    \defd{$\HB$}  & Hybrid with \mbox{Backward modalities}
                  & $φ \Coloneqq q \mid P \mid ¬φ \mid (φ_1∨φ_2) \mid \dm[R](φ_1,…,φ_k) \mid \bdm[R](φ_1,…,φ_k)$ \\\seplinespace
    \defd{$\HG$}  & Hybrid with \mbox{Global modalities}
                  & $φ \Coloneqq q \mid P \mid ¬φ \mid (φ_1∨φ_2) \mid \dm[R](φ_1,…,φ_k) \mid \gdm φ$ \\\seplinespace
    \defd{$\HS$}  & Hybrid with \mbox{Set quantifiers}
                  & $φ \Coloneqq q \mid P \mid ¬φ \mid (φ_1∨φ_2) \mid \dm[R](φ_1,…,φ_k) \mid \EE{P}φ$ \\\seplinespace
    \multicolumn{2}{l}{\defd{$\HBG$},\, \defd{$\HBS$},\, \defd{$\HGS$} \,and\, \defd{$\HBGS$}}
                  & analogous to the preceding grammars \\\seplinespace \addlinespace
    \defd{$\FO$}  & \mbox{First-Order}
                  & $φ \Coloneqq (p\eqsymb q) \mid P(q) \mid R(q_0,…,q_k) \mid ¬φ \mid (φ_1∨φ_2) \mid \EE{q}φ$ \\\seplinespace
    \defd{$\MSO$} & Monadic \mbox{Second-Order}
                  & $φ \Coloneqq (p\eqsymb q) \mid P(q) \mid R(q_0,…,q_k) \mid ¬φ \mid (φ_1∨φ_2) \mid \EE{q}φ \mid \EE{P}φ$ \\
    \addlinespace\midrule
    \multicolumn{3}{l}{Here,\, $p,q,q_0,…,q_k∈\ElemSyms$,\: $P∈\SetSyms$, and $R∈\RelSyms[k+1]$, for $k≥1$.} \\
    \bottomrule
  \end{tabularx}
  \caption{Languages of the considered logics.}
  \label{tab:languages}
\end{table*}

\Cref{tab:syntax-semantics} shows how formulas are built up, and what they mean.
Furthermore,
it indicates how to obtain
the set \defd{$\free{φ}$} of symbols that occur freely in a given formula $φ$,
i.e., outside the scope of a binding operator.
If $\free{φ} ⊆ \Signature$,
we say that $φ$ is a \defd{sentence} over $\Signature$.
The relation $⊨$ defined in \cref{tab:syntax-semantics}
specifies in which cases a structure $\A$ satisfies $φ$,
written \defd{${\A⊨φ}$},
assuming that $φ$ is a sentence over $\signature{\A}$.
Otherwise, we stipulate that ${\A⊭φ}$.

Of particular interest for this paper are those formulas
in which the element symbol $\at$ is considered to be free,
although it might not occur explicitly.
They are evaluated on a pointed structure $\A$
from the perspective of the element $\inp{\A}{\at}$\!.
Atomic formulas of the form $q$ or $P$, with $q∈\ElemSyms$ and $P∈\SetSyms$,
are satisfied if $\inp{\A}{\at}$ is labeled by the corresponding symbol.
Using the operator~$\dm[R]$\!,
which is called the \mbox{\defd{$R$-diamond}},
we can remap the symbol $\at$ through existential quantification
over the elements in $\A$ that are reachable from $\inp{\A}{\at}$
through the relation $\inp{\A}{R}$.
If we want to do the same with respect to the inverse relation of $\inp{\A}{R}$,
we can use the \defd{backward $R$-diamond}~$\bdm[R]$.
In addition,
there is also the \defd{global diamond}~$\gdm$
(unfortunately often called “universal modality”),
which ranges over all elements of $\A$.
It can be considered as the diamond operator corresponding
to the edge relation of the complete graph over $\domain{\A}$.
To facilitate certain descriptions,
we shall sometimes treat $\bdm[R]$ and $\gdm$ as (extended) special cases of $\dm[R]$,
assuming that they are implicitly associated
with the reserved relation symbols \defd{$\invR$} and \defd{$\Tglob$}, respectively.
These symbols do not belong to $\Symbols$,
and therefore cannot be interpreted by any structure.

Allowing a bit of syntactic sugar,
we will make liberal use of the remaining operators of predicate logic,
i.e., $∧$, $→$, $↔$, $∀$,
and we may leave out some parentheses,
assuming that~$∨$ and $∧$ take precedence over $→$ and~$↔$.
Furthermore,
we define the abbreviations
\begin{align*}
  &⊤ \defeq \at, \quad ⊥ \defeq ¬\at \quad \text{and} \\
  &\bx[R](φ_1,…,φ_k) \,\defeq\, ¬\dm[R](¬φ_1,…,¬φ_k).
\end{align*}
Note that the first line makes sense because, by definition,
the atomic formula $\at$ is always satisfied at the point of evaluation.
Also,
the second line remains applicable if one substitutes $\invR$ or $\Tglob$ for $R$.
The resulting operators $\bx[R]$, $\bbx[R]$ and $\gbx$
provide universal quantification
and are called \defd{boxes}
(using the same attributes as for diamonds).
Diamonds and boxes are collectively referred to as \mbox{\defd{modalities}} or \defd{modal operators}.
In case we restrict ourselves to structures that only have a single relation,
we may omit the relation symbol $R$, and just use empty modalities such as~$\dm$.
Similarly,
if the relation symbols involved are indexed, like $R_1,…,R_u$,
we associate them with modalities of the form~$\dm[i]$, for $1≤i≤u$.

Let us now turn to the specific classes of formulas
considered in this article,
which are presented in \cref{tab:languages}.
The first-order ($\FO$) and monadic second-order ($\MSO$) languages
are defined in the usual way.
The remaining classes can all be qualified as modal languages,
insofar as they include modal operators,
but not the classical first-order quantifiers.
We refer to them as \mbox{\defd{hybrid}} languages
because, unlike basic modal logic, they also provide element symbols
(or “nominals”).
This is consistent with the terminology of the modal logic community
(see, e.g., \cite{AC06} or \cite{BRV02}).
However,
there does not seem to be any established alphabetical nomenclature
that is both concise and easily extensible to fit our purposes.
Therefore,
we introduce the following system:
Starting with the letter H, for “hybrid”,
we add B or G
if we want to include backward or global modalities, respectively.
In the same manner,
the letter S gives us
the key ingredient investigated in this paper~---
namely, set quantifiers.
With this,
the classes SOPML and SOPMLE of \cite{Kuu15} become $\HS$ and~$\HGS$.

For any set of formulas $Φ$ (e.g., $\HGS$),
we shall refer to its members as \defd{$Φ$-formulas}.
Given such a $Φ$-formula $φ$
and a class of structures $\C$ (e.g., $\DIGRAPH$),
we use the semantic bracket notations $\sem[\C]{φ}$ and $\sem[\C]{Φ}$
to denote the set of structures defined by $φ$ over $\C$,
and the family of sets definable in $Φ$ over $\C$.
More formally,
\newcommand*{\StrutSemCPhi}{\vphantom{\sem[\C]{φ}}}
\begin{align*}
  \defd{\swl{\sem[\C]{φ}}{\sem[\C]{Φ}}} &\defeq \LRsetbuilder{\A∈\C}{\A⊨φ \StrutSemCPhi},
  \quad \text{and} \\
  \defd{\sem[\C]{Φ}} &\defeq \LRsetbuilder{\sem[\C]{φ}}{φ∈Φ}.
\end{align*}
If $\C$ is equal to the set of all structures,
we omit the subscript and simply write \defd{$\sem{φ}$} and \defd{$\sem{Φ}$}.
Similarly,
we use ${\defd{\eqcl[\C]{φ}} \defeq \LRsetbuilder{ψ}{\sem[\C]{ψ}=\sem[\C]{φ}}}$
for the equivalence class of $φ$ over $\C$,
and $\defd{\eqcl[\C]{Φ}} \defeq \bigcup_{φ∈Φ}\eqcl[\C]{φ}$
for the set of all formulas that are equivalent over $\C$ to some formula in $Φ$.
We may again drop the subscript
if we do not want to restrict to a particular class of structures.

\subsection{A Useful Example}
\label{ssec:singleton}
As we do not allow first-order quantification in our hybrid formulas,
some properties that seem very natural in $\FO$
become rather cumbersome to express.
Nevertheless,
translation from $\FO$ to $\HGS$ is always possible
because we can simulate first-order quantifiers
by set quantifiers relativized to singletons,
which, by extension, also leads to
the equivalence between $\MSO$ and $\HGS$.
With this in mind,
let us consider the following formula schema,
where $X∈\SetSyms$,\, $R∈\RelSyms[2]$,
and $φ$ can be any $\HBG$-formula:
\begin{equation*}
  \defd{\seeone[R](φ)} \defeq
  \dm[R] φ \,∧\;
  \AA{X} \bigl( \dm[R](φ {\,∧\,} X) → \bx[R](φ {\,→\,} X) \bigr).
\end{equation*}
When evaluated on a pointed structure $\A$
whose signature includes $\set{\at,R}∪\free{φ}$,
the formula $\seeone[R](φ)$ states that there is exactly one element $a∈\A$
reachable from $\inp{\A}{\at}$ through an $\inp{\A}{R}$-edge,
such that $φ$ is satisfied at $a$ (i.e., by the structure $\ver{\A}{\at}{a}$).
In the context of $1$-relational graphs,
we may use the shorthand \defd{$\seeone(φ)$} to invoke this schema.
Coming back to the original motivation,
we also~define
\begin{equation*}
  \defd{\totone(φ)} \defeq \seeone[\Tglob](φ),
\end{equation*}
which states that
there is precisely one element in the entire structure $\A$
at which $φ$ is satisfied.
Here,
$\A$ does not necessarily have to be pointed,
and, of course, $\signature{\A}$ never contains $\Tglob$.

Anticipating the notation of the next subsection,
the formulas obtained by this construction
can be classified as~$\eqcl{\PS{1}(Φ)}$-formulas,
where $Φ∈\set{\H,\HB,\HG,\HBG}$
depends on the specific modalities that we use.

\subsection{Alternation Hierarchies}
\label{ssec:alternation}
We now come to our primary objects of interest.
Assume we are given some set of formulas $Φ$,
referred to as \defd{kernel},
which is free of set quantifiers and closed under negation
(e.g., $\HG$).
Then,
for $ℓ≥0$, the class \defd{$\SS{ℓ}(Φ)$} consists of those formulas
that one can construct by taking a member of $Φ$
and prepending to it at most~$ℓ$ consecutive blocks of set quantifiers,
alternating between existential and universal blocks,
such that the first block is existential.
Reformulating this
solely in terms of existential quantifiers and negations,
we get
\begin{align*}
  \SS{0}(Φ)   &\defeq Φ \quad \text{and} \\
  \SS{ℓ+1}(Φ) &\defeq \LRsetbuilder{\EE{P}}{P∈\SetSyms}^*\!·\LRsetbuilder{¬φ}{φ∈\SS{ℓ}(Φ)},
\end{align*}
where the second line uses set concatenation and the Kleene star.
We define \defd{$\PS{ℓ}(Φ)$} as the corresponding dual class,
i.e., the set of all negations of formulas in $\SS{ℓ}(Φ)$.
Generalizing this to arbitrary Boolean combinations,
let \defd{$\BC\SS{ℓ}(Φ)$} denote
the smallest superclass of $\SS{ℓ}(Φ)$
that is closed under negation and disjunction.

The formulas in $\SS{ℓ}(Φ)$ and $\PS{ℓ}(Φ)$
are said to be in \defd{prenex normal form}
with respect to the kernel $Φ$.
It is well known that every $\MSO$-formula
can be transformed into prenex normal form
with kernel class $\FO$.
This is based on the observation that first-order quantifiers
can be replaced by second-order ones.
Using the construction of \cref{ssec:singleton},
it is not difficult to see that
the analogue holds for $\HS$, $\HBS$, $\HGS$ and $\HBGS$
with respect to their corresponding kernel classes.
A more elaborate explanation can be found in \cite[Prp.~3]{Cat06}.

For the sake of clarity,
we break with the tradition of implicit quantification
that is customary in modal logic.
Instead of evaluating $\HS$-formulas on non-pointed structures
by means of “hidden” universal quantification,
we shall explicitly put a global box in front of our formulas.
This leads to the class
\begin{equation*}
  \defd{\gbx\SS{ℓ}(\H)} \defeq \set{\gbx}·\SS{ℓ}(\H).
\end{equation*}
Analogously, we also define $\defd{\gbx\PS{ℓ}(\H)}$.

All of our results will be stated in terms of the semantic classes
that one obtains by evaluating the preceding formula classes
on some set of structures $\C$.
On the semantic side,
we will additionally consider the class
\begin{equation*}
  \defd{\sem[\C]{\DS{ℓ}(Φ)}} \defeq \sem[\C]{\SS{ℓ}(Φ)} ∩ \sem[\C]{\PS{ℓ}(Φ)}.
\end{equation*}
Since it is not based on any syntactic counterpart,
there is no meaning attributed to
the notation $\DS{ℓ}(Φ)$ by itself (without the brackets).

%% file: tex/results.tex
\section{Main Results}
\label{sec:results}

\begin{table*}[tp]
  \centering
  \newcommand*{\seplinespace}{\addlinespace[2\defaultaddspace]}
  \begin{tabular}{lllrl@{\hspace{4ex}}l}
    \toprule
    \textit{Separation result} & \textit{Kernel} & \textit{Structures} & \textit{Levels}
                                                                       & \multicolumn{2}{l}{\textit{Theorem}} \\
        & Class $Φ$ & Class $\C$ & $ℓ≥{·}$\, \\
    \midrule\addlinespace
    $\sem[\C]{\DS{ℓ+1}(Φ)}⊈\sem[\C]{\BC\SS{ℓ}(Φ)}$
        & $\FO$         & $\GRID$, $\DIGRAPH$, $\GRAPH$ & $1$ && \ttheorem{MST}{thm:DB-FO} $\bast$ \\
        & $\HBG$, $\HG$ & $\GRID$, $\DIGRAPH$, $\GRAPH[1,1]$ & $1$ && \ttheorem{R}{thm:DB-HBG-HG} \\\seplinespace
    $\sem[\C]{\SS{ℓ}(Φ)}\incomparable\sem[\C]{\PS{ℓ}(Φ)}$
        & $\FO$         & $\GRID$, $\DIGRAPH$, $\GRAPH$ & $1$ && \ttheorem{MST}{thm:SP-FO} $\bast$ \\
        & $\HBG$, $\HG$ & $\GRID$, $\DIGRAPH$, $\GRAPH[1,1]$ & $1$ && \ttheorem{R}{thm:SP-HBG-HG} \\
        & $\H$          & $\PDIGRAPH$ & $1$ && \ttheorem{R}{thm:SP-H} \\\seplinespace
    $\sem[\C]{\gbx\SS{ℓ}(Φ)}⊈\sem[\C]{\gbx\PS{ℓ}(Φ)}$
        & $\H$          & $\DIGRAPH$ & $2$ && \ttheorem{R}{thm:gSP-H} \\
    \addlinespace[1.5\defaultaddspace]\bottomrule
  \end{tabular}
  \captionsetup{justification=centering}
  \caption{The specific separation results of \cref{thm:separation-MST,thm:separation-R}. \\
    \Cref{thm:separation-MST} (marked by asterisks for better visibility)
    is due to Matz, Schweikardt and Thomas.}
  \label{tab:results}
\end{table*}

With the notation in place,
we are ready to formally enunciate the main \lcnamecref{thm:separation-R},
whose complete proof will be the subject of the remainder of this paper.
It is an extension to modal kernel formulas
of the following result of Matz, Schweikardt and Thomas,
which can be obtained by combining
\mbox{\cite[Thm.~1]{MST02}} and \mbox{\cite[Thm.~2.26]{Mat02}}\footnote{
  \cite[Thm.~2.26]{Mat02} states that
  $\sem[\GRID]{\SS{ℓ}(\FO)}⊉\sem[\GRID]{\PS{ℓ}(\FO)}$,
  which, by duality, also implies
  $\sem[\GRID]{\SS{ℓ}(\FO)}⊈\sem[\GRID]{\PS{ℓ}(\FO)}$.
}:

\begin{theorem}[Matz, Schweikardt, Thomas]
  \label{thm:separation-MST}
  The set quantifier alternation hierarchy of $\MSO$ is strict
  over the classes of grids, directed graphs and undirected graphs.

  \emph{A more precise statement of this theorem, referred to as
    \cref{thm:separation-MST}~\ref{thm:DB-FO}~and~\ref{thm:SP-FO},
    is given in \cref{tab:results}.}
\end{theorem}

Roughly speaking,
the extension provided in the present paper tells us
that the preceding separations are largely maintained
if we replace the first-order kernel by certain classes of modal formulas.
To facilitate comparisons,
the formal statements of both \lcnamecrefs{thm:separation-MST}
are presented together in the same \lcnamecref{tab:results}.

\begin{theorem}[Main Results]
  \label{thm:separation-R}
  The set quantifier alternation hierarchies of $\HBGS$ and $\HGS$ are strict
  over the classes of grids, directed graphs and $1$-bit labeled undirected graphs.

  Furthermore, the corresponding hierarchies of $\HS$ and~$\gbx\+\HS$ are (mostly) strict
  over the classes of pointed directed graphs and directed graphs, respectively.

  \emph{A more precise statement of this theorem, referred to as
    \cref{thm:separation-R}~\ref{thm:DB-HBG-HG},~\ref{thm:SP-HBG-HG},~\ref{thm:SP-H}~and~\ref{thm:gSP-H},
    is given in \cref{tab:results}.}
\end{theorem}

\begin{figure}[tp]
  \centering
  \input{fig/hierarchy-nodelta.tex}
  \caption{The set quantifier alternation hierarchy implied by
    \cref{thm:separation-R}~\ref{thm:gSP-H}\,
    for $Φ = \H$,\, $\C = \DIGRAPH$\, and $ℓ≥2$.}
  \label{fig:hierarchy-nodelta}
\end{figure}
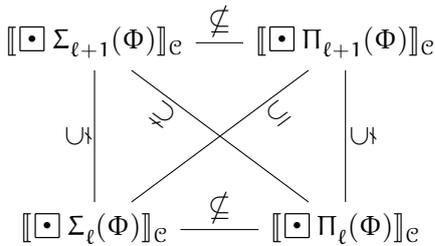

By basic properties of predicate logic\footnote{
  In particular, the inclusion
  $\sem[\C]{\BC\SS{ℓ}(Φ)} ⊆ \sem[\C]{\DS{ℓ+1}(Φ)}$
  follows from the fact that,
  when transforming a Boolean combination of $\SS{ℓ}(Φ)$-formulas
  into prenex normal form,
  one is free to choose whether the resulting formula
  (with up to $ℓ+1$ quantifier alternations)
  should start with an existential or a universal quantifier.}
and the transitivity of set inclusion,
it is easy to infer from \cref{thm:separation-R}
the hierarchy diagrams represented in
\cref{fig:hierarchy-delta,fig:hierarchy-nodelta}.

If we take into account all the depicted relations,
the diagram in \cref{fig:hierarchy-delta}
is the same as in \cite{MST02} and \cite{Mat02}.
Hence,
when switching to one of the modal kernels that include global modalities,
i.e., $\HBG$ or $\HG$,
the separations of \cref{thm:separation-MST}
are completely preserved on grids and directed graphs.
Our proof method also allows us to
easily transfer this result to undirected graphs,
as long as we admit that
the vertices may be labeled with at least one bit.
Additional work would be required to eliminate this condition.

\begin{figure}[tp]
  \centering
  \input{fig/hierarchy-delta.tex}
  \caption{The set quantifier alternation hierarchies established by 
    \cref{thm:separation-R}~\ref{thm:DB-HBG-HG},~\ref{thm:SP-HBG-HG}~and~\ref{thm:SP-H}.
    If we include the noninclusion in parentheses,
    this diagram holds for
    $Φ{\:∈\,}\set{\HBG,\HG}$\, and\, $\C{\:∈\,}\set{\GRID,\DIGRAPH,\GRAPH[1,1]}$.
    If we ignore that noninclusion,
    it is also verified for
    $Φ = \H$\, and\, $\C = \PDIGRAPH$.
    In both cases, we assume $ℓ≥1$.}
  \label{fig:hierarchy-delta}
\end{figure}
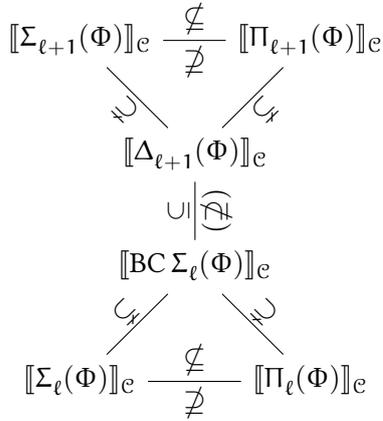

As a spin-off,
\cref{thm:separation-R} also provides an extension
of some of these separations to $\H$, 
a kernel class without global modalities.
Following \cite{Kuu08,Kuu15},
we consider the alternation hierarchies of both $\HS$ and $\gbx\+\HS$.
For the former,
which is evaluated on pointed directed graphs,
\cref{fig:hierarchy-delta} gives a detailed picture,
leaving open only whether the inclusion
$\sem[\C]{\BC\SS{ℓ}(Φ)} ⊆ \sem[\C]{\DS{ℓ+1}(Φ)}$ is proper.
Inferring the strictness of this inclusion from the preceding results
does not seem very difficult,
but would call for a generalization of our framework.
In contrast,
the second hierarchy based on $\H$ is arguably less natural,
since every $\gbx\+\HS$-formula is prefixed by a global box,
regardless of the occurring set quantifiers.
This creates a certain asymmetry
between the $\SS{ℓ}$-{} and $\PS{ℓ}$-levels,
which becomes apparent when considering
the missing relations in \cref{fig:hierarchy-nodelta}.
Unlike for the other hierarchies,
one cannot simply argue by duality
to deduce from
$\sem[\C]{\gbx\SS{ℓ}(Φ)}⊈\sem[\C]{\gbx\PS{ℓ}(Φ)}$
that the converse noninclusion also holds.
Nevertheless,
the presented result is strong enough to completely settle
the specific strictness question mentioned in~\cite{Kuu08},
and left unanswered in~\cite{Kuu15}:
For arbitrarily high~$ℓ$, we have
\begin{equation*}
  \sem[\DIGRAPH]{\gbx\SS{ℓ}(\H)}⊉\sem[\DIGRAPH]{\gbx\SS{ℓ+1}(\H)} \+.
\end{equation*}

%% file: fig/hierarchy-delta.tex
\begin{tikzpicture}[node distance=13ex]
  \node (b1) {$\sem[\C]{\BC\SS{ℓ}(Φ)}$};
  \node (s1) [below left of=b1] {$\sem[\C]{\SS{ℓ}(Φ)}$};
  \node (p1) [below right of=b1] {$\sem[\C]{\PS{ℓ}(Φ)}$};
  \node (d2) [above of=b1,yshift=-4ex] {$\sem[\C]{\DS{ℓ+1}(Φ)}$};
  \node (s2) [above left of=d2] {$\sem[\C]{\SS{ℓ+1}(Φ)}$};
  \node (p2) [above right of=d2] {$\sem[\C]{\PS{ℓ+1}(Φ)}$};
  \path[every node/.style={sloped,allow upside down,auto=false,inner sep=0.4ex}]
    (s2) edge node [anchor=south] {$⊈$}
              node [anchor=north] {$⊉$} (p2)
    (s2) edge node [anchor=north] {$\supsetneq$} (d2)
    (d2) edge node [anchor=north] {$\subsetneq$} (p2)
    (b1) edge node [anchor=south] {$⊆$}
              node [anchor=north] {$(⊉)$} (d2)
    (s1) edge node [anchor=south] {$\subsetneq$} (b1)
    (b1) edge node [anchor=south] {$\supsetneq$} (p1)
    (s1) edge node [anchor=south] {$⊈$}
              node [anchor=north] {$⊉$} (p1);
\end{tikzpicture}

%% file: fig/hierarchy-nodelta.tex
\begin{tikzpicture}[node distance=15ex]
  \node (s1) {$\sem[\C]{\gbx\SS{ℓ}(Φ)}$};
  \node (p1) [right of=s1,xshift=5ex] {$\sem[\C]{\gbx\PS{ℓ}(Φ)}$};
  \node (s2) [above of=s1] {$\sem[\C]{\gbx\SS{ℓ+1}(Φ)}$};
  \node (p2) [above of=p1] {$\sem[\C]{\gbx\PS{ℓ+1}(Φ)}$};
  \newlength{\labelshift}\setlength{\labelshift}{9ex}
  \path[every node/.style={sloped,allow upside down,auto=false,inner sep=0.4ex}]
    (s2) edge node [anchor=south] {$⊈$} (p2)
    (s1) edge node [anchor=south] {$\subsetneq$} (s2)
    (s2) edge node [anchor=north] {\hspace{-1.1\labelshift} $\supsetneq$} (p1)
    (s1) edge node [anchor=north] {\hspace{\labelshift} $⊆$} (p2)
    (p1) edge node [anchor=north] {$\subsetneq$}(p2)
    (s1) edge node [anchor=south] {$⊈$} (p1);
\end{tikzpicture}

%% file: tex/proofs.tex
\section{Top-Level Proofs}
\label{sec:proofs}
In accordance with our top-down approach,
the present \lcnamecref{sec:proofs} already provides the proof
of our main \lcnamecref{thm:separation-R},
where everything comes together.
It therefore acts as a gateway to the \lcnamecrefs{sec:grids}
with the technical parts,
especially \cref{sec:encodings}.

\subsection{Figurative Inclusions}
First of all,
we need to introduce the primary tool with which
we will transfer separation results from one setting to another.
It can be seen as an abstraction of the \emph{strong first-order reductions}
used in \cite{MST02}.
Unlike the latter,
it is formulated independently of any logical language,
which allows us to postpone the technical details to the end of the paper.

\begin{definition}[Figurative Inclusion]
  \NoMathBreak
  Consider two sets $\C$ and $\D$
  and a partial \emph{injective} function $μ\colon\C\pto\D$.
  For any two families of subsets $\L⊆2^\C$ and $\M⊆2^\D$\!,
  we say that $\L$ is \defd{forward included} in $\M$
  \defd{figuro} $μ$, and write \defd{$\L \figsubeq{μ} \M$},
  if for every set $L∈\L$, there is a set $M∈\M$
  such that $μ(L) = M ∩ μ(\C)$.
\end{definition}

Figuratively speaking,
the partial bijection $μ$ creates a tunnel between $\C$ and $\D$,
and all the sets in $\L$ and $\M$ are cropped
to fit through that tunnel.
Two original sets are considered to be equal
if their cropped versions are mapped onto each other by~$μ$.

We also define the shorthands $\figsupeq{μ}$\, and $\figeq{μ}$\,
as natural extensions of the previous notation:
\defd{$\L\figsupeq{μ}\M$},
which is defined as $\M\figsubeq{\inv{μ}}\L$,
means that $\M$ is \defd{backward included} in $\L$ \defd{figuro} $μ$,
and \defd{$\L\figeq{μ}\M$},
an abbreviation for the conjunction of
$\L\figsubeq{μ}\M$ and $\L\figsupeq{μ}\M$,
states that $\L$ is \defd{forward equal} to $\M$ \defd{figuro} $μ$.
All of these relations are referred to as \defd{figurative inclusions}.

Note that ordinary inclusion is
a special case of figurative inclusion, i.e.,
for $\C = \D$,
\begin{equation*}
  \L ⊆ \M \quad \text{\Iff} \quad \L \figsubeq{\id[\C]} \M.
\end{equation*}
Furthermore, figurative inclusion is transitive in the sense that
\begin{equation*}
  \L \,\figsubeq{μ}\, \M \,\figsubeq{ν}\, \N
  \quad \text{implies} \quad
  \L \,\figsubeq{ν∘μ}\, \N.
\end{equation*}
(This depends crucially on the fact that $ν$ is injective.)
% \begin{proof}
%   Consider three sets $\C$, $\D$ and $\E$,
%   two partial injective functions $μ\colon\C\pto\D$ and $ν\colon\D\pto\E$,
%   and three families of subsets $\L⊆2^\C$,\, $\M⊆2^\D$ and $\N⊆2^\E$.
%   Assume that we have $\L \,\figsubeq{μ}\, \M \,\figsubeq{ν}\, \N$.
%   Choose an arbitrary set $L∈\L$.
%   Since $\L \figsubeq{μ} \M$,
%   there must be a set $M∈\M$ such that $μ(L) = M ∩ μ(\C)$.
%   Furthermore, as $\M \figsubeq{ν} \N$,
%   there is also a set $N∈\N$ such that $ν(M) = N ∩ ν(\D)$.
%   Hence,
%   \begin{align*}
%     (ν∘μ)(L) &= ν(M ∩ μ(\C)) \\
%              &= ν(M) ∩ (ν∘μ)(\C) \tag{$*$}\label{eq:distributivity} \\
%              &= N ∩ ν(\D) ∩ (ν∘μ)(\C) \\
%              &= N ∩ (ν∘μ)(\C).
%   \end{align*}
%   Equality~\eqref{eq:distributivity} holds because $ν$ is injective.
%   Since the choice of $L$ was arbitrary,
%   there is such an $N∈\N$ for every $L∈\L$,
%   and thus $\L \figsubeq{ν∘μ} \N$.
% \end{proof}

In our specific context,
given a noninclusion $\sem[\C]{Φ_2} ⊈ \sem[\C]{Φ_1}$,
we shall use the concept of figurative inclusion to infer from it
another noninclusion $\sem[\D]{Ψ_2} ⊈ \sem[\D]{Ψ_1}$.
Here,
$Φ_1,Φ_2,Ψ_1,Ψ_2$ and $\C,\D$
refer to some classes of formulas and structures, respectively.
The key part of the argument will be
to construct an appropriate encoding function $μ\colon\C\to\D$,
in order to apply the following \lcnamecref{lem:separation-transfer}.~

\begin{lemma}
  \label{lem:separation-transfer}
  \NoMathBreak
  Let $\L_1,\L_2⊆2^\C$ and $\M_1,\M_2⊆2^\D$ be families of subsets of some sets $\C$ and $\D$.
  If there is a \emph{total \mbox{injective}} function $μ\colon\C\to\D$
  such that $\L_2 \figsubeq{μ} \M_2$ and $\L_1 \figsupeq{μ} \M_1$,
  then
  \begin{equation*}
    \L_2 ⊈ \L_1 \quad \text{implies} \quad \M_2 ⊈ \M_1. \qedhere
  \end{equation*}
\end{lemma}
\begin{proof}
  \NoMathBreak
  To show the contrapositive,
  let us suppose that $\M_2 ⊆ \M_1$,
  or, equivalently, $\M_2 \figsubeq{\id[\D]} \M_1$.
  Then the chain of figurative inclusions
  \begin{equation*}
    \L_2 \:\figsubeq{μ}\: \M_2 \:\figsubeq{\id[\D]}\: \M_1 \:\figsubeq{\inv{μ}}\: \L_1
  \end{equation*}
  yields $\L_2 \figsubeq{\id[\C]} \L_1$,
  since $(\inv{μ}\!∘\id[\D]∘μ) = \id[\C]$.
  (This depends on $μ$ being total and injective.)
  Consequently, we have $\L_2 ⊆ \L_1$.
\end{proof}

In some cases,
we can combine two given figurative inclusions in order to obtain a new one
that relates the corresponding intersection classes.
This property will be very useful for establishing figurative inclusions
between classes of the form $\sem[\C]{\DS{ℓ}(Φ)}$.

\begin{lemma}
  \label{lem:figsubeq-intersection}
  \NoMathBreak
  Consider two sets $\C$ and $\D$,
  a partial injective function $μ\colon\C\pto\D$,
  and four families of subsets $\L_1,\L_2⊆2^\C$ and $\M_1,\M_2⊆2^\D$.
  If $μ(\C)$ is a member of $\M_1∩\M_2$,
  and $\M_1$, $\M_2$ are both closed under intersection, then
  \begin{gather*}
   \L_1 \figsubeq{μ} \M_1 \quad \text{and} \quad \L_2 \figsubeq{μ} \M_2 \\
   \quad \text{imply} \quad \L_1∩\L_2 \,\figsubeq{μ}\, \M_1∩\M_2.
   \qedhere
  \end{gather*}
\end{lemma}
\begin{proof}
  Let $L$ be any set in $\L_1∩\L_2$.
  Since $\L_1 \figsubeq{μ} \M_1$,
  there is, by definition, a set $M$ in $\M_1$
  such that $μ(L) = M ∩ μ(\C)$.
  Furthermore, we also know that $μ(\C)$ lies in $\M_1$,
  and that the latter is closed under intersection.
  Hence, $μ(L)∈\M_1$.
  Analogously, we also get that $μ(L)∈\M_2$.
  Finally, knowing that
  for all $L$ in $\L_1∩\L_2$,
  $μ(L)$ lies in $\M_1∩\M_2$,
  we obviously have a sufficient condition for
  $\L_1∩\L_2 \figsubeq{μ} \M_1∩\M_2$.
\end{proof}

\subsection{Proving the Main Theorem}
\label{ssec:main-proof}
We are now ready to give the central proof of this paper.
Although it makes references to many statements of \cref{sec:grids,sec:encodings},
it is formulated in a way that can be understood without having read
anything beyond this point.

\begin{proof}[Proof of \cref{thm:separation-R}]
  The basis of our proof shall be laid in \cref{sec:grids},
  where the case $t=0$ of \cref{thm:levelwise-equivalence}
  will state the following:
  When restricted to the class of grids,
  the set quantifier alternation hierarchies of
  $\MSO$, $\HBGS$ and $\HGS$ are equivalent.
  More precisely,
  for every $ℓ≥1$ and
  $Ξ ∈ \set{\SS{ℓ},\,\PS{ℓ},\,\BC\SS{ℓ},\,\DS{ℓ}}$,
  it holds that
  \begin{equation*}
    \sem[\GRID]{\+Ξ\+(\FO)} \:=\; \sem[\GRID]{\+Ξ\+(\HBG)} \:=\; \sem[\GRID]{\+Ξ\+(\HG)} \+.
  \end{equation*}
  Hence,
  if we consider only the case $\C=\GRID$,
  the separation results for the kernel class $\FO$ stated in
  \cref{thm:separation-MST}~\ref{thm:DB-FO}~and~\ref{thm:SP-FO}
  immediately imply those for $\HBG$ and $\HG$ in
  \cref{thm:separation-R}~\ref{thm:DB-HBG-HG}~and~\ref{thm:SP-HBG-HG}.

  The remainder of the proof now consists of
  establishing suitable figurative inclusions,
  in order to transfer these results
  to other classes of structures and,
  to some extent,
  to weaker classes of kernel formulas.
  For this purpose,
  we shall introduce in \cref{sec:encodings} a notion of translatability
  between two classes of kernel formulas $Φ$ and $Ψ$,
  with respect to a given total injective function $μ$
  that encodes structures from a class~$\C$ into structures of some class~$\D$.
  As will be shown in \cref{lem:translation-inclusion},
  bidirectional translatability implies
  \begin{equation}
    \label{eqn:levelwise-figeq}
    \sem[\C]{\+Ξ\+(Φ)} \;\figeq{μ}\; \sem[\D]{\+Ξ\+(Ψ)}
    \tag{$\ast$}
  \end{equation}
  for all $Ξ ∈ \set{\SS{ℓ},\,\PS{ℓ},\,\BC\SS{ℓ}}$ with $ℓ≥0$.
  If we can additionally show that $μ(\C)$ is (at most)
  $\DS{2}(Ψ)$-definable over $\D$,
  then, by \cref{lem:figsubeq-intersection},
  the figurative equality \eqref{eqn:levelwise-figeq}
  also holds for $Ξ = \DS{ℓ+1}$ with $ℓ≥1$.
  Note that the backward part “$\figsupeq{μ}$\!” is always true,
  since $\inv{μ}(\D)$ is trivially $\DS{2}(Φ)$-definable over $\C$.

  The groundwork being in place,
  we proceed by applying \cref{lem:separation-transfer} as follows:
  \begin{itemize}[nosep]
  \item If we have established \eqref{eqn:levelwise-figeq} for $Ξ ∈ \set{\SS{ℓ},\,\PS{ℓ}}$,
    then we can transfer the separation
    \begin{equation}
      \label{eqn:separation-SP}
      \sem[\C]{\SS{ℓ}(Φ)}\incomparable\sem[\C]{\PS{ℓ}(Φ)}
      \tag{1}
    \end{equation}
    to the kernel class $Ψ$ evaluated on the class of structures~$\D$.
  \item Similarly,
    if \eqref{eqn:levelwise-figeq} holds
    for $Ξ ∈ \set{\BC\SS{ℓ},\,\DS{ℓ+1}}$,
    then
    \begin{equation}
      \label{eqn:separation-DB}
      \sem[\C]{\DS{ℓ+1}(Φ)}⊈\sem[\C]{\BC\SS{ℓ}(Φ)}
      \tag{2}
    \end{equation}
    can also be transferred to $Ψ$ on $\D$.
  \end{itemize}

  It remains to provide concrete figurative inclusions
  to prove the different parts of \cref{thm:separation-R}.

  \proofparagraph{\ref{thm:DB-HBG-HG}\:\!,~\ref{thm:SP-HBG-HG}}
  The first two parts are treated in parallel.
  We start by transferring \eqref{eqn:separation-SP} and \eqref{eqn:separation-DB}
  from grids to directed graphs, for the kernel class $\HBG$,
  taking a detour via $2$-relational directed graphs,
  and, subsequently, via $2$-bit labeled ones.
  For all $Ξ ∈ \set{\SS{ℓ},\,\PS{ℓ},\,\BC\SS{ℓ},\,\DS{ℓ+1}}$ with $ℓ≥1$,
  we get
  \begin{alignat*}{2}
    \sem[\GRID]{\+Ξ\+(\HBG)} \;&\figeq{\id[\+\GRID]} & \;&\sem[{\DIGRAPH[0,2]}]{\+Ξ\+(\HBG)} \\
                               &\figeq{μ_1}          &   &\sem[{\DIGRAPH[2,1]}]{\+Ξ\+(\HBG)} \\
                               &\figeq{μ_2}          &   &\sem[\DIGRAPH]{\+Ξ\+(\HBG)} \+.
  \end{alignat*}
  The first line is trivial for $Ξ ∈ \set{\SS{ℓ},\,\PS{ℓ},\,\BC\SS{ℓ}}$,
  since $\GRID ⊆ \DIGRAPH[0,2]$.
  It also holds for $Ξ = \DS{ℓ+1}$ because
  $\GRID$ is $\PS{1}(\HBG)$-definable over $\DIGRAPH[0,2]$,
  as shall be demonstrated in \cref{prp:grid-definability}.
  The other two lines rely on the existence of
  adequate injective functions $μ_1$ and $μ_2$
  that allow us to apply
  \crefnosort{lem:translation-inclusion,lem:figsubeq-intersection}
  in the way explained above.
  They will be provided by
  \cref{prp:multirelational-labeled,prp:labeled-unlabeled},
  respectively.

  We proceed in a similar way
  to transfer \eqref{eqn:separation-SP} and \eqref{eqn:separation-DB}
  from $\HBG$ to $\HG$ on directed graphs:
  \begin{alignat*}{2}
    \sem[\DIGRAPH]{\+Ξ\+(\HBG)} \,&\figeq{μ_3} & \;\,&\sem[{\DIGRAPH[0,2]}]{\+Ξ\+(\HG)} \\
                                  &\figeq{μ_1} &     &\sem[{\DIGRAPH[2,1]}]{\+Ξ\+(\HG)} \\
                                  &\figeq{μ_2} &     &\sem[\DIGRAPH]{\+Ξ\+(\HG)} \+,
  \end{alignat*}
  for $Ξ ∈ \set{\SS{ℓ},\,\PS{ℓ},\,\BC\SS{ℓ},\,\DS{ℓ+1}}$ with $ℓ≥1$.
  The very simple encoding function $μ_3$,
  which lets us eliminate backward modalities
  and again use \crefnosort{lem:translation-inclusion,lem:figsubeq-intersection},
  will be supplied by \cref{prp:hbg-to-hg}.
  The encodings $μ_1$ and $μ_2$ are the same as before,
  because the properties asserted by
  \cref{prp:multirelational-labeled,prp:labeled-unlabeled} 
  hold for both $\HG$ and $\HBG$ as kernel classes.
  Incidentally, this means
  we could transfer \eqref{eqn:separation-SP}
  directly from $\HG$ on grids to $\HG$ on directed graphs,
  without even mentioning $\HBG$.

  To show that \eqref{eqn:separation-SP} and \eqref{eqn:separation-DB}
  are also valid for $\HG$ on $1$-bit labeled undirected graphs,
  we establish
  \begin{equation*}
    \sem[\DIGRAPH]{\+Ξ\+(\HBG)} \, \figeq{μ_4} \, \sem[{\GRAPH[1,1]}]{\+Ξ\+(\HG)} \+,
  \end{equation*}
  again for all $Ξ ∈ \set{\SS{ℓ},\,\PS{ℓ},\,\BC\SS{ℓ},\,\DS{ℓ+1}}$ with $ℓ≥1$.
  The appropriate encoding $μ_4$ shall be constructed in \cref{prp:digraph-1bitgraph}.
  Since backward modalities do not offer any additional expressive power
  on undirected graphs,
  the separations we obtain also hold for the kernel $\HBG$.

  \proofparagraph{\ref{thm:SP-H}}
  Next,
  to transfer \eqref{eqn:separation-SP}
  from $\HG$ on directed graphs to $\H$ on pointed directed graphs,
  we show that, for $Ξ ∈ \set{\SS{ℓ},\,\PS{ℓ}}$ with $ℓ≥1$,
  we have
  \begin{equation*}
    \sem[\DIGRAPH]{\+Ξ\+(\HG)} \:\figeq{μ_5}\; \sem[\PDIGRAPH]{\+Ξ\+(\H)} \+.
  \end{equation*}
  The injective function $μ_5$,
  which satisfies the translatability property
  required to obtain this figurative equality via \cref{lem:translation-inclusion},
  will be provided by \cref{prp:digraph-pdigraph}.
  Its image $μ_5(\DIGRAPH)$ is not $\HS$-definable,
  for the simple reason
  that an $\HS$-formula is unable to distinguish between two structures
  that are isomorphic when restricted to
  the connected component containing the position marker $\at$.
  Hence, we cannot merely apply \cref{lem:figsubeq-intersection}
  to show~\eqref{eqn:separation-DB}.
  Our approach would have to be refined
  to take into account equivalence classes of structures,
  which we shall not do in this paper.

  \proofparagraph{\ref{thm:gSP-H}}
  Finally,
  \cref{prp:digraph-pdigraph} will also state that
  $μ_5$ can be converted into an encoding $μ_5'$, from $\DIGRAPH$ back into $\DIGRAPH$,
  that satisfies the following figurative inclusions for all $ℓ≥2$:
  \begin{alignat*}{2}
    \sem[\DIGRAPH]{\SS{ℓ}(\HG)} &\,\figsubeq{μ_5'}\;\, & &\sem[\DIGRAPH]{\gbx\SS{ℓ}(\H)} \,, \\
    \sem[\DIGRAPH]{\PS{ℓ}(\HG)} &\,\figeq{μ_5'}\;\,    & &\sem[\DIGRAPH]{\gbx\PS{ℓ}(\H)} \,.
  \end{alignat*}
  Using \eqref{eqn:separation-SP} for $\HG$ on directed graphs,
  and applying \cref{lem:separation-transfer},
  we can infer from this that
  \begin{equation*}
    \sem[\DIGRAPH]{\gbx\SS{ℓ}(\H)}⊈\sem[\DIGRAPH]{\gbx\PS{ℓ}(\H)} \+.
    \qedhere
  \end{equation*}
\end{proof}

%% file: tex/grids.tex
\section{Grids as a Starting Point}
\label{sec:grids}
In this \lcnamecref{sec:grids},
we establish that the set quantifier alternation hierarchies of
$\MSO$, $\HBGS$ and $\HGS$ are equivalent on labeled grids.
In addition,
we give a $\eqcl{\PS{1}(\HBG)}$-formula that characterizes the class of grids.

\subsection{The Standard Translation}
Our first building block is a well-known property of modal logic,
which holds even if we do not confine ourselves to the setting of grids.

\begin{proposition}
  \label{prp:standard-translation}
  For every $\HBG$-formula, there is an equivalent $\FO$-formula, i.e., 
  \begin{equation*}
    \sem{\HBG} \;⊆\; \sem{\FO} \+.
    \qedhere
  \end{equation*}
\end{proposition}
\begin{proof}
  Given an $\HBG$-formula $φ$,
  we have to construct an $\FO$-formula $ψ_φ$
  such that $\A⊨φ$ \Iff{} $\A⊨ψ_φ$,
  for every structure $\A$.
  This is simply a matter of transcribing the semantics of $\HBG$
  given in \cref{tab:syntax-semantics}
  to the language of first-order logic,
  a method known as the \emph{standard translation} in modal logic
  (see, e.g., \cite[Def.~2.45]{BRV02}).
  The following table gives a recursive specification of this translation.
  \\[\abovedisplayskip]
  \begin{tabularx}{\linewidth}{lY}
    \toprule
    $φ∈\HBG$             & Equivalent formula\, $ψ_φ\!∈\FO$ \\
    \midrule
    $q$                  & $\at \eqsymb q$ \\\addlinespace
    $Q$                  & $Q(\at)$ \\\addlinespace
    $¬φ_1$               & $¬ψ_{φ_1}$ \\\addlinespace
    $φ_1 ∨ φ_2$          & $ψ_{φ_1} ∨ ψ_{φ_2}$ \\\addlinespace
    $\dm[R](φ_1,…,φ_k)$  & $\EE{x_1,…,x_k}\bigl(\,
                            R(\at,x_1,…,x_k) \;∧{}$
                            \hspace*{\fill} $\bigwedge_{1≤i≤k}\ver{ψ_{φ_i}}{\at}{x_i}
                            \,\bigr)$ \\\addlinespace
    $\bdm[R](φ_1,…,φ_k)$ & \mbox{as above, except $R(x_k,…,x_1,\at)$} \\\addlinespace
    $\gdm φ_1$           & $\EE{\at}\+ ψ_{φ_1}$ \\\addlinespace
    \bottomrule
  \end{tabularx} \\[\belowdisplayskip]
  Here, $q∈\ElemSyms$,\; $Q∈\SetSyms$,\; $R∈\RelSyms[k+1]$,\;
  $φ_1,…,φ_k∈\HBG$, for $k≥1$,
  and $x_1,…,x_k$ are element symbols,
  chosen such that $x_i ∉ \free{ψ_{φ_i}}$.
  The notation $\ver{ψ_{φ_i}}{\at}{x_i}$ designates the formula obtained by
  substituting each free occurrence of $\at$ in $ψ_{φ_i}$ by $x_i$.
\end{proof}

\subsection{A Detour through Tiling Systems}
By restricting our focus to the class of labeled grids,
we can take advantage of a well-studied automaton model
introduced by Giammarresi and Restivo in \cite{GR92},
which is closely related to $\MSO$.
A “machine” in this model,
called a \defd{tiling system},
is defined as a tuple $\T = \tuple{Γ,Ω,Θ}$,\, where
\begin{itemize}
\item $Γ = \set{0,1}^t$ is seen as an alphabet, with $t≥0$,
\item $Ω$ is a finite set of sates,\, and
\item $Θ ⊆ \bigl((Γ×Ω)∪\set{\hash}\bigr)^{4\mathstrut}$
  is a set of $2{×}2$-tiles
  that may use a fresh symbol $\hash$ not contained in $(Γ×Ω)$.
  \qedhere
\end{itemize}
For a fixed number of bits $t$,
we denote by \defd{$\TS_t$} the set of all tiling systems with alphabet $Γ = \set{0,1}^t$.

Given a $t$-bit labeled grid $\sC$,
a tiling system $\T∈\TS_t$ operates similarly to a nondeterministic finite automaton 
generalized to two dimensions.
A run of $\T$ on $\sC$ is an extended labeled grid $\sC^\hash$\!,
obtained by nondeterministically labeling each cell of $\sC$
with some state $ω∈Ω$
and surrounding the entire grid with a border
consisting of new $\hash$-labeled cells.
We consider $\sC^\hash$ to be a valid run
if each of its $2{×}2$-subgrids can be identified with some tile in $Θ$.
The set recognized by $\T$ 
consists precisely of those labeled grids
for which such a run exists.
By analogy with our existing notation,
we write \defd{$\sem{\TS_t}$}
for the class formed by
the sets of $t$-bit labeled grids that are recognized by some tiling system in $\TS_t$.

Exploiting a locality property of first-order logic,
Giammarresi, Restivo, Seibert and Thomas
have shown in \cite{GRST96}
that tiling systems capture precisely
the existential fragment of $\MSO$ on labeled grids:

\begin{theorem}[Giammarresi, Restivo, Seibert, Thomas]
  \label{thm:equivalence-ts-emso}
  For arbitrary $t≥0$, a set of $t$-bit labeled grids
  is \mbox{$\TS$-recognizable} \Iff{} it is $\SS{1}(\FO)$-definable
  over $\GRID[t]$, i.e.,
  \begin{equation*}
    \sem{\TS_t} \;=\; \sem[{\GRID[t]}]{\SS{1}(\FO)} \+.
    \qedhere
  \end{equation*}
\end{theorem}

The preceding result is extremely useful for our purposes,
because, from the perspective of modal logic,
it provides a much easier access to $\MSO$.
This brings us to the following \lcnamecref{prp:inclusion-ts-ehgs}.

\begin{proposition}
  \label{prp:inclusion-ts-ehgs}
  For arbitrary $t≥0$, if a set of $t$-bit labeled grids
  is \mbox{$\TS$-recognizable}, then it is also $\SS{1}(\HG)$-definable
  over $\GRID[t]$, i.e., 
  \begin{equation*}
    \sem{\TS_t} \;⊆\; \sem[{\GRID[t]}]{\SS{1}(\HG)} \+.
    \qedhere
  \end{equation*}
\end{proposition}
\begin{proof}
  \newcommand*{\Middle}{{\operatorname{M}}}
  \newcommand*{\Top}{{\operatorname{T}}}
  \newcommand*{\Bottom}{{\operatorname{B}}}
  \newcommand*{\Left}{{\operatorname{L}}}
  \newcommand*{\Right}{{\operatorname{R}}}
  \newcommand*{\border}{{\operatorname{border}}}
  Let $\T = \tuple{Γ,Ω,Θ}$ be a tiling system
  with alphabet $Γ = \set{0,1}^t$.
  We have to construct a $\SS{1}(\HG)$-sentence $φ_\T$
  over the signature $\set{P_1,…,P_t,R_1,R_2}$,
  such that each labeled grid $\sC∈\GRID[t]$ satisfies $φ_\T$
  \Iff{} it is accepted by $\T$.

  The idea is standard:
  We represent the states of $\T$ by additional set symbols $\tuple{X_ω}_{ω∈Ω}$,
  and our formula asserts that there exists a corresponding partition of $\domain{\sC}$
  into $\card{Ω}$ subsets
  that represent a run $\sC^\hash$ of $\T$ on~$\sC$.
  To verify that it is indeed a valid run,
  we have to check that each $2{×}2$-subgrid of $\sC^\hash$
  corresponds to some tile
  \begin{equation*}
    θ = 
    \begin{bmatrix*}[l]
      θ_1 & θ_2 \\
      θ_3 & θ_4
    \end{bmatrix*}
  \end{equation*}
  in $Θ$.
  If the entry $θ_1$ is different from $\hash$,
  we can easily write down an $\H$-formula $φ_θ$
  that checks at a given position $c∈\sC$,
  whether the $2{×}2$-subgrid of $\sC^\hash$ with upper-left corner $c$
  matches $θ$.
  Here, $θ_1$ is chosen as the representative entry of $θ$,
  because the upper-left corner of the tile can “see” the other elements
  by following the directed $R_1$-{} and $R_2$-edges.
  Otherwise, if $θ_1$ is equal to $\hash$, there is no such element $c$,
  since $\sC$ does not contain special border elements.
  However, we can always choose some other entry $θ_i$, different from $\hash$,
  to be the representative of $θ$,
  and write a formula $φ_θ$ describing the tile
  from the point of view of an element corresponding to $θ_i$.
  This choice is never arbitrary,
  because the representative must be able to “see”
  the other \mbox{non-$\hash$} entries of the tile.
  Consequently,
  we divide $Θ$ into four disjoint sets $Θ_1$, $Θ_2$, $Θ_3$, $Θ_4$,
  such that $Θ_i$ contains those tiles $θ$ that are represented by their entry $θ_i$.
  In order to facilitate the subsequent formalization,
  we further subdivide each set into partitions
  according to the $\hash$-borders that occur within the tiles:
  $Θ_\Middle$ contains the “middle tiles” (all entries different from~$\hash$),
  $Θ_\Left$ the “left tiles” (with $θ_1$ and $θ_3$ equal to~$\hash$),
  $Θ_{\Bottom\Right}$ the “bottom-right tiles”,
  and so forth~…
  Altogether,
  $Θ$ is partitioned into nine subsets, grouped into four types:
  \begin{alignat*}{2}
    Θ_1 &= Θ_\Middle \dcup Θ_\Bottom \dcup Θ_\Right \dcup Θ_{\Bottom\Right} \qquad& Θ_2 &= Θ_\Left \dcup Θ_{\Bottom\Left} \\
    Θ_3 &= Θ_\Top \dcup Θ_{\Top\Right}                                       & Θ_4 &= Θ_{\Top\Left}
  \end{alignat*}

  We now construct the formula $φ_\T$ in a bottom-up manner,
  starting with a subformula $φ_{θ_i}$
  for each entry $θ_i$ other than $\hash$, for every tile $θ∈Θ$.
  Letting $θ_i$ be equal to $\tuple{γ,ω}∈Γ×Ω$, with $γ = \tuple{γ_j}_{1≤j≤t}$,
  the formula $φ_{θ_i}$ checks at a given position $c∈\sC$
  if the labeling of $c$ matches~$θ_i$.
  \begin{equation*}
    φ_{θ_i} \+=\; \smashoperator{\bigwedge_{γ_j=1}}P_j \,∧\,
                 \smashoperator{\bigwedge_{γ_j=0}}¬P_j \,∧\,
                 X_ω \,∧\,
                 \smashoperator{\bigwedge_{ω'≠ω}}¬X_{ω'}
  \end{equation*}

  Building on this, we can define for each tile $θ∈Θ$
  the formula $φ_θ$ mentioned above.
  Since $\HG$ does not have backward modalities,
  there is a certain asymmetry between tiles in $Θ_1$,
  where the representative can “see” the entire $2{×}2$-subgrid,
  and the remaining tiles,
  where the representative must “know”
  that it lies in the leftmost column or the uppermost row of the grid $\sC$.
  We shall address this issue shortly,
  and just assume that information not accessible to the representative
  is verified by another part of the ultimate formula $φ_\T$.
  For tiles in $Θ_\Middle$, $Θ_{\Bottom\Right}$, $Θ_\Left$, $Θ_{\Top\Left}$,
  the definitions of $φ_θ$ are given in the following table.
  For tiles in $Θ_\Bottom$, $Θ_\Right$, $Θ_{\Bottom\Left}$, $Θ_\Top$, $Θ_{\Top\Right}$,
  the method is completely analogous. \\[\abovedisplayskip]
  \begin{tabularx}{\linewidth}{rY}
    \toprule
    $\swl{\;θ}{
      \bigl[ \begin{smallmatrix*}[l]
        θ_1 & θ_2 \\
        θ_3 & θ_4
      \end{smallmatrix*} \bigr]
    }$
    & $φ_θ$ \\
    \midrule\addlinespace
    $\swl{Θ_\Middle}{Θ_{\Bottom\Right}} ∋
      \bigl[ \begin{smallmatrix*}[l]
        θ_1 & θ_2 \\
        θ_3 & θ_4
      \end{smallmatrix*} \bigr]$
    & $φ_{θ_1} {\+∧\,} \dm[2]φ_{θ_2} {\+∧\,} \dm[1]φ_{θ_3} {\+∧\,} \dm[1]\dm[2]φ_{θ_4}$ \\\addlinespace
    $Θ_{\Bottom\Right} ∋
      \bigl[ \begin{smallmatrix*}[l]
        θ_1 & \hash \\
        \hash & \swl{\scriptstyle\hash}{θ_4}
      \end{smallmatrix*} \bigr]$
    & $φ_{θ_1} ∧\, \bx[2]⊥ \+∧\, \bx[1]⊥$ \\\addlinespace
    $\swl{Θ_\Left}{Θ_{\Bottom\Right}} ∋
      \bigl[ \begin{smallmatrix*}[l]
        \hash & θ_2 \\
        \swl{\scriptstyle\hash}{θ_3} & θ_4
      \end{smallmatrix*} \bigr]$
    & $φ_{θ_2} ∧ \dm[1]φ_{θ_4}$ \\\addlinespace
    $Θ_{\Top\Left} ∋
      \bigl[ \begin{smallmatrix*}[l]
        \hash & \hash \\
        \swl{\scriptstyle\hash}{θ_3} & θ_4
      \end{smallmatrix*} \bigr]$
    & $φ_{θ_4}$ \\\addlinespace
    \bottomrule
  \end{tabularx} \vspace{\belowdisplayskip}

  It remains to mark the top and left borders of $\sC$,
  using two additional predicates $Y_\Top$ and $Y_\Left$,
  over which we will quantify existentially.
  To this end,
  we write an $\HG$-formula $φ_\border$, checking
  that top [resp.~left] elements
  have no $R_1$-{} [resp.~$R_2$-] predecessor,
  that there is a top-left element,
  and that being top [resp.~left] is passed on to the $R_2$-{} [resp.~$R_1$-] successor,
  if it exists.
  \begin{equation*}
    \begin{aligned}
      φ_\border \+=\; ¬&\gdm(\dm[1]Y_\Top∨\dm[2]Y_\Left) \;∧\; \gdm(Y_\Top∧Y_\Left) \;∧{} \\[-0.5ex]
                     &\+\+\gbx\Bigl((Y_\Top{\,→\,}\bx[2]Y_\Top) ∧ (Y_\Left{\,→\,}\bx[1]Y_\Left)\Bigr)
    \end{aligned}
  \end{equation*}

  Finally,
  we can put everything together
  to describe the acceptance condition of $\T$.
  Every element $c∈\sC$ has to ensure
  that it corresponds to the upper-left corner of some tile in $Θ_1$.
  Furthermore, elements in the leftmost column or uppermost row of $\sC$
  must additionally check that the assignment of states is compatible
  with the tiles in $Θ_2$, $Θ_3$, $Θ_4$.
  This leads to the desired formula~$φ_\T$:
  \begin{align*}
    &\EE{\tuple{X_ω}_{ω∈Ω},\,Y_\Top,\,Y_\Left}\biggl( φ_\border \;∧{} \\[-0.5ex]
    &\qquad \gbx\bigl(\,\smashoperator{\bigvee_{θ ∈ Θ_1}}\!φ_θ\bigr)
            \;∧\; \gbx\bigl(Y_\Left→\smashoperator{\bigvee_{θ ∈ Θ_2}}\!φ_θ\bigr) \;∧{} \\[-1.5ex]
    &\qquad \gbx\bigl(Y_\Top→\smashoperator{\bigvee_{θ ∈ Θ_3}}\!φ_θ\bigr)
            \;∧\; \gbx\bigl(Y_\Top∧Y_\Left→\smashoperator{\bigvee_{θ ∈ Θ_4}}\!φ_θ\bigr)\biggr)
  \end{align*}
  Note that we do not need a separate subformula to check that the
  interpretations of $\tuple{X_ω}_{ω∈Ω}$ form a partition of $\domain{\sC}$,
  since this is already done implicitly in the conjunct $\gbx(\bigvee_{\!θ ∈ Θ_1}\!φ_θ)$. 
\end{proof}

\subsection{Equivalent Hierarchies on Grids}
Now we have everything at hand
to prove the levelwise equivalence of
$\MSO$, $\HBGS$ and $\HGS$ on labeled grids.

\begin{theorem}
  \label{thm:levelwise-equivalence}
  Let $t≥0$,\: $ℓ≥1$ and
  $Ξ ∈ \set{\SS{ℓ},\,\PS{ℓ},\,\BC\SS{ℓ},\,\DS{ℓ}}$.
  When restricted to the class of $t$-bit labeled grids,
  $Ξ\+(\FO)$, $Ξ\+(\HBG)$ and $Ξ\+(\HG)$
  are equivalent,\, i.e.,
  \begin{align*}
    \sem[{\GRID[t]}]{\+Ξ\+(\FO)} \:&=\; \sem[{\GRID[t]}]{\+Ξ\+(\HBG)} \\
                                   &=\; \sem[{\GRID[t]}]{\+Ξ\+(\HG)} \+.
    \qedhere
  \end{align*}
\end{theorem}
\begin{proof}
  First,
  we show that the claim holds for the case $Ξ = \SS{1}$
  (with arbitrary $t≥0$).
  This can be seen from the following circular chain of inclusions:
  \begin{align}
    \sem[{\GRID[t]}]{\SS{1}(\HG)} \:&⊆\; \sem[{\GRID[t]}]{\SS{1}(\HBG)} \label{eq:ehgs-ehbgs} \tag{a} \\
                                    &⊆\; \sem[{\GRID[t]}]{\SS{1}(\FO)} \label{eq:ehbgs-emso} \tag{b} \\
                                    &⊆\; \sem{\TS_t} \label{eq:emso-ts} \tag{c} \\
                                    &⊆\; \sem[{\GRID[t]}]{\SS{1}(\HG)} \label{eq:ts-ehgs} \tag{d}
  \end{align}
  \begin{enumerate}[after=\vspace{\belowdisplayskip}]
  \item[(\ref{eq:ehgs-ehbgs})]
    The first inclusion follows from the fact that
    $\SS{1}(\HG)$ is a syntactic fragment of $\SS{1}(\HBG)$.
  \item[(\ref{eq:ehbgs-emso})]
    For the second inclusion,
    consider any $\SS{1}(\HBG)$-formula $\widehat{φ} = \EE{Q_1,…,Q_n}(φ)$,
    where $Q_1,…,Q_n$ are set symbols and $φ$ is an $\HBG$-formula.
    By \cref{prp:standard-translation},
    we can replace $φ$ in $\widehat{φ}$ by an equivalent $\FO$-formula $ψ_φ$.
    This results in the $\SS{1}(\FO)$-formula $ψ_{\widehat{φ}} = \EE{Q_1,…,Q_n}(ψ_φ)$,
    which is equivalent to $\widehat{φ}$ on arbitrary structures,
    and thus, in particular, on $t$-bit labeled grids.
  \item[(\ref{eq:emso-ts})]
    The translation from $\SS{1}(\FO)$ on labeled grids to tiling systems
    corresponds to the more challenging direction of \cref{thm:equivalence-ts-emso},
    which is the main result of \cite{GRST96}.
  \item[(\ref{eq:ts-ehgs})]
    The last inclusion is given by \cref{prp:inclusion-ts-ehgs}.
  \end{enumerate}

  The general version of the \lcnamecref{thm:levelwise-equivalence}
  can now be obtained by induction on $ℓ$.
  This is straightforward,
  because the classes
  $\PS{ℓ}(Φ)$,\, $\BC\SS{ℓ}(Φ)$ and $\SS{ℓ+1}(Φ)$
  are defined syntactically in terms of $\SS{ℓ}(Φ)$,
  for any set of kernel formulas $Φ$
  (see \cref{ssec:alternation}),
  and if the claim holds for $Ξ ∈ \set{\SS{ℓ},\,\PS{ℓ}}$,
  then it also holds for the intersection classes of the form $\sem[{\GRID[t]}]{\DS{ℓ}(Φ)}$.
\end{proof}

\subsection{A Logical Characterization of Grids}
We conclude this \lcnamecref{sec:grids} by showing
that a single layer of universal set quantifiers is enough
to describe grids in $\HBGS$.

\begin{proposition}
  \label{prp:grid-definability}
  The set of all grids is $\PS{1}(\HBG)$-definable
  over \mbox{$2$-relational} directed graphs, i.e., 
  \begin{equation*}
    \GRID \;∈\; \sem[{\DIGRAPH[0,\+2]}]{\PS{1}(\HBG)} \+.
    \qedhere
  \end{equation*}
\end{proposition}
\begin{proof}
  \newcommand*{\Source}{{\operatorname{src}}}
  In the course of this proof,
  we give a list of properties,
  items~\labelcref{itm:injective-function,itm:reach-sink,itm:unique-source,itm:horiz-sources-linked,itm:down-right-diagonal,itm:down-right-commute},
  which are obviously necessary
  for a $2$-relational directed graph $\sD$ to be a grid,
  and show how to express them as $\eqcl{\PS{1}(\HBG)}$-formulas.
  We argue that the conjunction of all of these properties
  also constitutes a sufficient condition for being a grid,
  which immediately provides us with the required formula,
  since $\sem{\PS{1}(\HBG)}$ is closed under intersection.
  \begin{enumerate}
  \item \label{itm:injective-function}
    For each relation symbol $R ∈ \set{R_1,R_2}$,
    every element has at most one $R$-predecessor and at most one $R$-successor;
    in other words,
    $\inp{\sD}{R_1}$ and $\inp{\sD}{R_2}$ are partial injective functions.
    \begin{equation*}
      \smashoperator[r]{\bigwedge_{R\,∈\,\set{R_1,\,\invr{R_1}\!,\,R_2,\,\invr{R_2}}}} \:
      \AA{X}\+\gbx \bigl( \dm[R]X → \bx[R]X \bigr)
    \end{equation*}
  \item \label{itm:reach-sink}
    Again considering each $R ∈ \set{R_1,R_2}$ separately,
    there is a directed $R$-path from every element to an $R$-sink,
    i.e., to some element without $R$-successor.
    \begin{equation*}
      \smashoperator[r]{\bigwedge_{R\,∈\,\set{R_1,\,R_2}}} \,
      \AA{X} \Bigl(
      \gdm X {\,∧\,} \gbx\bigl( X {\,→\,} \bx[R]X \bigr)
      → \gdm\bigl( X {\,∧\,} \bx[R]⊥ \bigr)
      \Bigr)
    \end{equation*}
  \end{enumerate}
  Taken together,
  properties~\labelcref{itm:injective-function,itm:reach-sink}
  state that $\inp{\sD}{R_1}$ and $\inp{\sD}{R_2}$
  each form a collection of directed, acyclic, pairwise vertex-disjoint paths.
  Let us refer to the first elements of those paths
  as $R_1$-{} and $R_2$-sources, respectively.
  \begin{enumerate}[resume*]
  \item \label{itm:unique-source}
    There is precisely one element that is both an $R_1$-{} and an $R_2$-source.
    \begin{equation*}
      \totone\bigl(\+ \bbx[1]⊥ ∧ \bbx[2]⊥ \bigr)
    \end{equation*}
    (Here, $\totone$ is the schema from \cref{ssec:singleton}.)
  \item \label{itm:horiz-sources-linked}
    The $R_1$-predecessors and $R_1$-successors of $R_2$-sources
    must be $R_2$-sources themselves.
    \begin{equation*}
      \gbx \bigl(\+ \bbx[2]⊥ \+→\+ \bbx[1]\bbx[2]⊥ ∧ \bx[1]\bbx[2]⊥ \bigr)
    \end{equation*}
  \end{enumerate}
  By adding~\labelcref{itm:unique-source,itm:horiz-sources-linked}
  to our list of conditions,
  we ensure that there is an $R_1$-path consisting precisely of the $R_2$-sources,
  thereby also forcing the graph $\sD$ to be connected.
  \begin{enumerate}[resume*]
  \item \label{itm:down-right-diagonal}
    If an element has both an $R_1$- and an $R_2$-successor,
    then it also has a descendant
    reachable by first taking an $R_1$-edge and then an $R_2$-edge.
    \begin{equation*}
      \gbx \bigl( \dm[1]⊤ ∧ \dm[2]⊤ \+→\+ \dm[1]\dm[2]⊤ \bigr)
    \end{equation*}
  \item \label{itm:down-right-commute}
    The relations $\inp{\sD}{R_1}$ and $\inp{\sD}{R_2}$ commute.
    This means that following an $R_1$-edge and then an $R_2$-edge
    leads to the same element as
    first taking an $R_2$-edge and then an $R_1$-edge.
    \begin{equation*}
      \AA{X}\+\gbx \bigl( \dm[1]\dm[2]X ↔ \dm[2]\dm[1]X \bigr)
    \end{equation*}
  \end{enumerate}
  Considered in conjunction with
  condition~\labelcref{itm:injective-function},
  there are only two ways to
  satisfy~\labelcref{itm:down-right-diagonal,itm:down-right-commute}
  from the point of view of
  two elements $d_1,d_2∈\sD$ that are connected by an $R_1$-edge
  from $d_1$ to $d_2$:
  either both elements are $R_2$-sinks,
  or they have $R_2$-successors $d_1'$ and $d_2'$, respectively,
  with an $R_1$-edge from $d_1'$ to $d_2'$.
  Moreover,
  $d_2'$ only possesses an $R_1$-successor if $d_2$ does.
  Now, imagine we start from the left border,
  i.e., from the $R_1$-path that consists of all the $R_2$-sources,
  which is provided by
  properties~\labelcref{itm:injective-function,itm:reach-sink,itm:unique-source,itm:horiz-sources-linked},
  and iteratively enforce the requirements just mentioned.
  Then, in doing so, we propagate the grid topology through the entire graph.
  More specifically,
  the additional requirements of~\labelcref{itm:down-right-diagonal,itm:down-right-commute}
  entail that all the $R_2$-paths have the same length,
  and that the elements lying at a fixed (horizontal) position of those $R_2$-paths
  constitute an independent $R_1$-path,
  ordered in the same way as their respective $R_2$-predecessors.
\end{proof}

%% file: tex/encodings.tex
\section{A Toolbox of Encodings}
\label{sec:encodings}
In this \lcnamecref{sec:encodings},
we provide all the encoding functions
used in the proof of \cref{thm:separation-R} (see \cref{ssec:main-proof}),
and show that they satisfy suitable translatability properties,
allowing us to establish the required figurative inclusions.
With a view to modularity and reusability,
some of our constructions are more general than needed.

From here on,
integer intervals of the form $\setbuilder{i}{m≤i≤n}$
will be denoted by \defd{$\range[m]{n}$}.
We may also use the shorthand $\defd{\range{n}} \defeq \range[1]{n}$,
and, by analogy with the Bourbaki notation for real intervals,
\defd{$\lrange[m]{n}$} indicates that we exclude the endpoint $m$.
Furthermore,
given a set of symbols $σ$,
the extension $σ∪\set{\at}$ will be abbreviated to~$\defd{σ_\at}$.

\subsection{Encodings that Allow for Translation}
We shall only consider encoding functions
that are linear in the following sense:

\begin{definition}[Linear Encoding]
  \label{def:linear-encoding}
  \NoMathBreak
  Let $\C$, $\D$ be two classes of structures,
  and $m$, $n$ be integers such that $1≤m≤n$.
  A \defd{linear encoding} from $\C$ into $\D$ with parameters $m$, $n$
  is a total injective function $μ\colon\C\to\D$
  that assigns to each structure $\A∈\C$ a structure $μ(\A)∈\D$,
  whose domain is composed of $m$ disjoint copies of the domain of $\A$
  and $n-m$ additional elements, i.e.,
  \begin{equation*}
    \domain{μ(\A)} = \bigl(\range[1]{m}×\domain{\A}\bigr) \;∪\; \lrange[m]{n}.
    \qedhere
  \end{equation*}
\end{definition}

Given such a linear encoding $μ$ and some $\HBG$-formula $φ$,
we want to be able to construct a new formula $ψ_φ$,
such that
evaluating $φ$ on $\C$ is equivalent to evaluating $ψ_φ$ on $μ(\C)$.
Conversely,
we also desire a way of constructing a formula $φ_ψ$
that is equivalent on $\C$ to a given formula $ψ$ on $μ(\C)$.
The following two \lcnamecrefs{def:forward-translation}
formalize this translatability property for both directions.
We then show in \cref{lem:translation-inclusion}
that they adequately capture our intended meaning.
Although the underlying idea is very simple,
the presentation is a bit lengthy
because we have to exhaustively cover the structure of $\HBG$-formulas.

\begin{definition}[Forward Translation]
  \label{def:forward-translation}
  \NoMathBreak
  \NoMathIndent
  Consider two classes of structures $\C$ and $\D$
  over signatures $σ$ and $τ$, respectively,
  two classes of formulas $Φ,Ψ∈\set{\H,\,\HB,\,\HG,\,\HBG}$,
  and a linear encoding $μ\colon\C\to\D$.
  We say that $μ$
  \defd{allows for forward translation from $Φ$ to $Ψ$}
  if the following properties are satisfied:
  \begin{enumerate}
  \item \label{itm:fwd-set}
    For each element symbol or set symbol $P$ in $σ$,
    there is a $Ψ$-sentence $ψ_P$ over $τ_\at$, such that 
    \begin{equation*}
      \ver{\A}{\at}{a} \,⊨\, P \quad\text{iff}\quad \ver{μ(\A)}{\at}{\tuple{1,a}} \,⊨\, ψ_P\+,
    \end{equation*}
    for all $\A∈\C$ and $a∈\A$.
  \item \label{itm:fwd-relation}
    For each relation symbol $R$ in $σ$ of arity $k+1≥2$,
    there is a $Ψ$-sentence $ψ_R$ over $τ_\at$
    enriched with additional set symbols $\tuple{Y_i}_{1≤i≤k}$, such that
    \begin{align*}
      &\bigver{\A}{\at,\tuple{X_i}_{i≤k}}{a,\tuple{A_i}_{i≤k}} \,⊨\, \dm[R]\tuple{X_i}_{i≤k} \\
      &\indentcont\text{\Iff} \\
      &\bigver{μ(\A)}{\at,\tuple{Y_i}_{i≤k}}{\tuple{1,a},\tuple{B_i}_{i≤k}} \,⊨\, ψ_R\+, \\[0.5ex]
      &\indentcont\text{assuming $A_i,B_i$ satisfy $a'{∈\,}A_i⇔\tuple{1,a'}{\,∈\,}B_i$,}
    \end{align*}
    for all $\A∈\C$,\; $a∈\A$, sets $\tuple{A_i}_{1≤i≤k}⊆\A$
    and $\tuple{B_i}_{1≤i≤k}⊆μ(\A)$,
    and set symbols $\tuple{X_i}_{1≤i≤k}$.
  \item \label{itm:fwd-backward}
    If $Φ$ includes backward modalities,
    then for each relation symbol $R$ in $σ$ of arity at least $2$,
    there is a $Ψ$-formula $ψ_\invR$ that satisfies
    the property of item~\ref{itm:fwd-relation} for $\invR$ instead of $R$.
  \item \label{itm:fwd-global}
    If $Φ$ includes global modalities,
    then there is a \mbox{$Ψ$-formula} $\psiglob$ that satisfies
    the property of item~\ref{itm:fwd-relation} for~$\Tglob$ instead of $R$ and $k=1$.
  \item \label{itm:fwd-initial}
    \newcommand*{\XgdmX}{\NoHeight{\scalebox{0.8}{$\displaystyle\frac{X}{\gdm X \vphantom{\big(}}$}}}
    There is a $Ψ$-sentence $ψ_\ini$ over $τ$
    enriched with an additional set symbol $Y$, such that
    \begin{align*}
      &\ver{\A}{X}{A} \,⊨\, \XgdmX
      \quad\text{iff}\quad
      \ver{μ(\A)}{Y}{B} \,⊨\, ψ_\ini, \\[1ex]
      &\indentcont\text{assuming $A,B$ satisfy $a∈A⇔\tuple{1,a}∈B$,} \\
      &\indentcont\text{where $\XgdmX$ is \,$X$\, if $\at∈σ$,\, and $\gdm X$ otherwise,}
    \end{align*}
    for all $\A∈\C$,\, $A⊆\A$,\, $B⊆μ(\A)$ and $X∈\RelSyms[1]$.
    \qedhere
  \end{enumerate}
\end{definition}

\pagebreak[3]
\begin{definition}[Backward Translation]
  \label{def:backward-translation}
  \NoMathBreak
  \NoMathIndent
  Consider two classes of structures $\C$ and $\D$
  over signatures $σ$ and $τ$, respectively,
  two classes of formulas $Φ,Ψ∈\set{\H,\,\HB,\,\HG,\,\HBG}$,
  and a linear encoding $μ\colon\C\to\D$
  with parameters $m$, $n$.
  We say that $μ$
  \defd{allows for backward translation from $Ψ$ to $Φ$}
  if the following properties are satisfied:
  \begin{enumerate}
  \item \label{itm:bwd-set}
    For each element symbol or set symbol $Q$ in $τ$ and all $h∈\range{n}$,
    there is a $Φ$-sentence $φ_Q^h$ over $σ_\at$, such that
    \begin{align*}
      &\ver{\A}{\at}{a} \,⊨\, φ_Q^h \quad\text{iff}\quad \ver{μ(\A)}{\at}{b} \,⊨\, Q, \\[0.5ex]
      &\indentcont\text{where $b$ is $\tuple{h,a}$ if $h≤m$, and $h$ otherwise,}
    \end{align*}
    for all $\A∈\C$ and $a∈\A$.
  \item \label{itm:bwd-relation}
    For each relation symbol $S$ in $τ$ of arity $k+1≥2$, and all $h∈\range{n}$,
    there is a $Φ$-sentence $φ_S^h$ over $σ_\at$ \strut
    enriched with additional set symbols $\tuple{X_i^j}_{1≤i≤k}^{1≤j≤n}$,\, such that
    \begin{align*}
      &\bigver{\A}{\at,\tuple{X_i^j}_{i≤k}^{j≤n}}{a,\tuple{A_i^j}_{i≤k}^{j≤n}} \,⊨\, φ_S^h \\
      &\indentcont\text{\Iff} \\
      &\bigver{μ(\A)}{\at,\tuple{Y_i}_{i≤k}}{b, \tuple{B_i}_{i≤k}} \,⊨\, \dm[S]\tuple{Y_i}_{i≤k}, \\[1ex]
      &\indentcont\text{where $b$ is $\tuple{h,a}$ if $h≤m$, otherwise $h$,\, and} \\
      &\indentcont B_i =\; \smashoperator{\bigcup_{1≤j≤m}}\,\bigl(\set{j}{×}A_i^j\bigr)
      \,∪\, \,\smashoperator{\bigcup_{m<j≤n}}\, \setbuilder{j}{A_i^j=\domain{\A}},
    \end{align*}
    for all structures $\A∈\C$, elements $a∈\A$, sets $\tuple{A_i^j}_{1≤i≤k}^{1≤j≤m}⊆\A$\,
    and $\tuple{A_i^j}_{1≤i≤k}^{m<j≤n}∈\set{∅,\domain{\A}}$,
    and set symbols $\tuple{Y_i}_{1≤i≤k}$.
  \item \label{itm:bwd-backward}
    If $Ψ$ includes backward modalities,
    then for each relation symbol $S$ in $τ$ of arity at least $2$, and all $h∈\range{n}$,
    there is a $Φ$-formula $φ_\invS^h$ that satisfies
    the property of item~\ref{itm:bwd-relation} for $\invS$ instead of $S$.
  \item \label{itm:bwd-global}
    If $Ψ$ includes global modalities, then for all $h∈\range{n}$,
    there is a $Φ$-formula $\phiglob[h]$ that satisfies
    the property of item~\ref{itm:bwd-relation} for $\Tglob$ instead of $S$ and $k=1$.
  \item \label{itm:bwd-initial}
    \newcommand*{\YgdmY}{\NoHeight{\scalebox{0.8}{$\displaystyle\frac{Y}{\gdm Y \vphantom{\big(}}$}}}
    There is a $Φ$-sentence $φ_\ini$ over $σ$
    enriched with additional set symbols $\tuple{X^j}^{1≤j≤n}$, such that
    \begin{align*}
      &\bigver{\A}{\tuple{X^j}^{j≤n}}{\tuple{A^j}^{j≤n}} \,⊨\, φ_\ini \\
      &\indentcont\text{\Iff} \\
      &\ver{μ(\A)}{Y}{B} \,⊨\, \YgdmY \,, \\[1ex]
      &\indentcont\text{where $\YgdmY$ is \,$Y$\, if $\at∈τ$,\, otherwise $\gdm Y$, and} \\[0.5ex]
      &\indentcont B =\; \smashoperator{\bigcup_{1≤j≤m}}\,\bigl(\set{j}{×}A^j\bigr)
      \,∪\, \,\smashoperator{\bigcup_{m<j≤n}}\, \setbuilder{j}{A^j=\domain{\A}},
    \end{align*}
    for all structures $\A∈\C$,\, sets $\tuple{A^j}^{1≤j≤m}⊆\A$
    and $\tuple{A^j}^{m<j≤n}∈\set{∅,\domain{\A}}$,
    and $Y∈\RelSyms[1]$.
    \qedhere
  \end{enumerate}
\end{definition}

To simplify matters slightly,
we shall say that a linear encoding $μ$
\defd{allows for bidirectional translation between $Φ$ and $Ψ$},
if it allows for both
forward  translation from $Φ$ to $Ψ$ and
backward translation from $Ψ$ to $Φ$.
Furthermore,
in case $Φ=Ψ$,
we may say “within $Φ$” instead of “between $Φ$ and $Φ$”.

Let us now prove
that our notion of translatability is indeed sufficient
to imply figurative inclusion on the semantic side,
even if we bring set quantifiers into~play.

\begin{lemma}
  \label{lem:translation-inclusion}
  Consider two classes of structures $\C$ and $\D$,
  a linear encoding $μ\colon\C\to\D$,
  two classes of formulas $Φ,Ψ∈\set{\H,\,\HB,\,\HG,\,\HBG}$,
  and let $Ξ ∈ \set{\SS{ℓ},\,\PS{ℓ},\,\BC\SS{ℓ}}$,\,
  for some arbitrary $ℓ≥0$.
  \begin{enumerate}
  \item \label{itm:forward-inclusion}
    If $μ$ allows for forward translation from $Φ$ to $Ψ$,
    then we have
    \begin{equation*}
      \sem[\C]{\+Ξ\+(Φ)} \;\figsubeq{μ}\; \sem[\D]{\+Ξ\+(Ψ)}\+.
    \end{equation*}
  \item \label{itm:backward-inclusion}
    Similarly,
    if $μ$ allows for backward translation from $Ψ$ to $Φ$,
    then we have
    \begin{equation*}
      \sem[\C]{\+Ξ\+(Φ)} \;\figsupeq{μ}\; \sem[\D]{\+Ξ\+(Ψ)}\+.
      \qedhere
    \end{equation*}
  \end{enumerate}
\end{lemma}
\begin{proof}
  Let $σ$ and $τ$ be the signatures underlying $\C$ and $\D$, respectively.
  Parts \ref{itm:forward-inclusion} and \ref{itm:backward-inclusion}
  of the \lcnamecref{lem:translation-inclusion}
  are treated separately in the following proof.

  In several places,
  given some $\HBGS$-formula $φ$,
  the need will arise to
  substitute newly created $\HBG$-formulas $φ_1,…,φ_k$ for set symbols $X_1,…,X_k$.
  We shall write \defd{$\ver{φ}{\tuple{X_i}_{i≤k}}{\tuple{φ_i}_{i≤k}}$}
  to denote the $\HBGS$-formula
  that one obtains by simultaneously replacing
  every free occurrence of each $X_i$ in $φ$ by the formula~$φ_i$.
  
  \proofparagraph{\ref{itm:forward-inclusion}}
  For every $Ξ\+(Φ)$-sentence $φ$ over $σ$,
  we must construct a $Ξ\+(Ψ)$-sentence $ψ_φ$ over $τ$,
  such that $ψ_φ$ says about $μ(\A)$ the same as $φ$ says about $\A$,
  for all structures $\A∈\C$.

  We start by focusing on the kernel classes $Φ,Ψ$,
  and show the following by induction on the structure of \mbox{$Φ$-formulas}:
  For every $Φ$-sentence $φ$ over \mbox{$σ_\at∪\Z$},
  with $\Z=\set{Z_1,…,Z_t}$ being any collection of set symbols disjoint from $σ$ and $τ$
  (i.e., “free set variables”),
  there is a $Ψ$-sentence $ψ_φ^*$ over $τ_\at∪\Z$ such that
  \begin{align*}
    &\bigver{\A}{\at,\tuple{Z_r}_{r≤t}}{a,\tuple{A_r}_{r≤t}} \,⊨\, φ \\
    &\quad\text{\Iff} \\
    &\bigver{μ(\A)}{\at,\tuple{Z_r}_{r≤t}}{\tuple{1,a},\tuple{B_r}_{r≤t}} \,⊨\, ψ_φ^*\+, \\[0.5ex]
    &\quad\text{assuming $A_r,B_r$ satisfy $a'{∈\,}A_r⇔\tuple{1,a'}{\,∈\,}B_r$,}
  \end{align*}
  for all structures $\A∈\C$, elements $a∈\A$,
  and sets $\tuple{A_r}_{1≤r≤t}⊆\A$ and $\tuple{B_r}_{1≤r≤t}⊆μ(\A)$.
  \begin{itemize}
  \item If $φ = \at$ or $φ = Z$, for some $Z∈\Z$,
    it suffices to set $ψ_φ^* = φ$.
  \item If $φ = P$, for some element symbol or set symbol $P$ in $σ$,
    we exploit that $μ$ allows for forward translation from $Φ$ to $Ψ$,
    and choose $ψ_φ^* = ψ_P$.
    Here, $ψ_P$ is the formula postulated by
    \cref{def:forward-translation}~\ref{itm:fwd-set};
    it fulfills the induction hypothesis,
    since adding interpretations of the symbols $Z_1,…,Z_t$
    to a structure
    has no influence on whether or not that structure satisfies
    a sentence over a signature that does not contain these symbols.
  \item If $φ = ¬φ_1$ \,or\, $φ = φ_1 ∨ φ_2$,
    where $φ_1$ and $φ_2$ are formulas that satisfy the induction hypothesis,
    we set $ψ_φ^* = ¬ψ_{φ_1}^*$ \,or\, $ψ_φ^* = ψ_{φ_1}^* ∨ ψ_{φ_2}^*$,\, respectively.
  \item If $φ = \dm[R]\tuple{φ_i}_{i≤k}$,
    where $R$ is a relation symbol in $σ$ of arity $k+1≥2$,
    and $\tuple{φ_i}_{i≤k}$ are $Φ$-sentences over $σ_\at∪\Z$
    satisfying the induction hypothesis,
    we again use the fact that $μ$ allows for forward translation from $Φ$ to $Ψ$.
    The desired formula $ψ_φ^*$ is obtained by substituting $\tuple{ψ_{φ_i}^*}_{i≤k}$
    for the symbols $\tuple{Y_i}_{i≤k}$ in the formula $ψ_R$,
    whose existence is asserted by
    \cref{def:forward-translation}~\ref{itm:fwd-relation}, i.e.,
    \begin{equation*}
      ψ_φ^* = \bigver{ψ_R}{\tuple{Y_i}_{i≤k}}{\tuple{ψ_{φ_i}^*}_{i≤k}}.
    \end{equation*}
    For $1≤i≤k$, let $A'_i$ be the set of elements $a'∈\A$
    that satisfy $φ_i$ in $\bigver{\A}{\tuple{Z_r}_{r≤t}}{\tuple{A_r}_{r≤t}}$,
    and let $B'_i$ be the set of elements $b'∈μ(\A)$
    that satisfy $ψ_{φ_i}$ in $\bigver{μ(\A)}{\tuple{Z_r}_{r≤t}}{\tuple{B_r}_{r≤t}}$.
    By induction hypothesis, we are guaranteed that
    all the sets $A'_i$, $B'_i$ are such that
    an element $a'$ lies in $A'_i$ \Iff{} $\tuple{1,a'}$ lies in $B'_i$.
    Thus, we have
    \begin{equation*}
      \begin{aligned}
        &\bigver{\A}{\at,\tuple{Z_r}_{r≤t}}{a,\tuple{A_r}_{r≤t}} \,⊨\, φ \\[0.5ex]
        \text{iff}\quad
        &\bigver{\A}{\at,\tuple{X_i}_{i≤k}}{a,\tuple{A'_i}_{i≤k}} \,⊨\, \dm[R]\tuple{X_i}_{i≤k} \\[0.5ex]
        \text{iff}\quad
        &\bigver{μ(\A)}{\at,\tuple{Y_i}_{i≤k}\:\!}{\tuple{1,a},\tuple{B'_i}_{i≤k}} \,⊨\, ψ_R \\[0.5ex]
        \text{iff}\quad
        &\bigver{μ(\A)}{\at,\tuple{Z_r}_{r≤t}}{\tuple{1,a},\tuple{B_r}_{r≤t}} \,⊨\, ψ_φ^*\+.
      \end{aligned}
    \end{equation*}
  \item If $φ = \bdm[R]\tuple{φ_i}_{i≤k}$,
    assuming $Φ$ incorporates backward modalities,
    we obtain $ψ_φ^*$ by applying the same argument as in the previous case,
    but this time considering $\invR$ instead of $R$ and invoking
    \cref{def:forward-translation}~\ref{itm:fwd-backward}.
  \item If $φ = \gdm φ_1$,
    supposing $Φ$ includes global modalities,
    we again follow the same line of reasoning as in the case $φ = \dm[R]\tuple{φ_i}_{i≤k}$,
    referring to \cref{def:forward-translation}~\ref{itm:fwd-global}
    and using~$\Tglob$ instead of $R$, with $k=1$.
  \end{itemize}

  Now we can consider the case where
  the position symbol~$\at$ is not (re)mapped,
  and then look beyond the kernel classes
  to finally deal with set quantifiers.
  Arguing once more by structural induction,
  we extend the preceding claim as follows:
  For every $Ξ\+(Φ)$-sentence $φ$ over $σ∪\Z$,
  with $\Z=\set{Z_1,…,Z_t}$ as before (possibly empty),
  there is a $Ξ\+(Ψ)$-sentence $ψ_φ$ over $τ∪\Z$ such that
  \begin{align*}
    &\bigver{\A}{\tuple{Z_r}_{r≤t}}{\tuple{A_r}_{r≤t}} \,⊨\, φ \\
    &\quad\text{\Iff} \\
    &\bigver{μ(\A)}{\tuple{Z_r}_{r≤t}}{\tuple{B_r}_{r≤t}} \,⊨\, ψ_φ\+, \\[0.5ex]
    &\quad\text{assuming $A_r,B_r$ satisfy $a{\,∈\,}A_r⇔\tuple{1,a}{\,∈\,}B_r$,}
  \end{align*}
  for all $\A∈\C$,\,
  $\tuple{A_r}_{1≤r≤t}⊆\A$ and $\tuple{B_r}_{1≤r≤t}⊆μ(\A)$.
  \begin{itemize}
  \item If $φ$ lies in the kernel $Φ$,
    we make use of the claim just proven,
    together with the formula $ψ_\ini$ described in
    \cref{def:forward-translation}~\ref{itm:fwd-initial}.
    We set\, $ψ_φ = \ver{ψ_\ini}{Y}{ψ_φ^*}$.
    \begin{itemize}
    \item If $\at$ belongs to $σ$,
      the asserted property of $ψ_\ini$ guarantees
      that $φ$ holds at the initial position $\inp{\A}{\at}$
      in the $\Z$-extended variant of $\A$
      \Iff{} $ψ_φ$ is satisfied by the $\Z$-extended variant of $μ(\A)$.
    \item Otherwise, $\at$ cannot be free in $φ$,
      since $φ$ is a sentence over $σ∪\Z$,
      which also implies that $Φ$ incorporates global modalities.
      It follows that $φ$ is equivalent to $\gdm φ$.
      Again applying the definition of $ψ_\ini$,
      we obtain that the $\Z$-extended variant of $\A$ satisfies $\gdm φ$,
      and thus $φ$,
      \Iff{} the $\Z$-extended variant of $μ(\A)$ satisfies $ψ_φ$.
    \end{itemize}
  \item If $φ$ is a Boolean combination of formulas
    that satisfy the induction hypothesis,
    the translation is straightforward,
    just as in the previous part of the proof.
  \item If $φ = \EE{Z_{t+1}}φ_1$,
    where $φ_1$ is a $Ξ\+(Φ)$-sentence over \mbox{$σ∪\set{Z_1,…,Z_{t+1}}$}
    that satisfies the hypothesis,
    we choose $ψ_φ = \EE{Z_{t+1}}ψ_{φ_1}$.
    To justify this choice,
    let $\A'$ and $μ(\A)'$ denote the $\Z$-extended variants of $\A$ and $μ(\A)$, respectively.
    We get the following by induction:
    \begin{itemize}
    \item If choosing $\map{Z_{t+1}}{A_{t+1}}$ leads to satisfaction of $φ_1$ in $\A'$,
      then choosing $\map{Z_{t+1}}{\set{1}\+{×}\+A_{t+1}}$ does the same for $ψ_{φ_1}$ in $μ(\A)'$.
    \item Conversely, if $\map{Z_{t+1}}{B_{t+1}}$ is a satisfying choice for $ψ_{φ_1}$ in $μ(\A)'$,
      then so is $\map{Z_{t+1}}{{\setbuilder{a}{\tuple{1,a}∈B_{t+1}}}}$ for $φ_1$ in $\A'$.
    \end{itemize}
  \end{itemize}

  \proofparagraph{\ref{itm:backward-inclusion}}
  The proof of the reverse direction of the \lcnamecref{lem:translation-inclusion}
  is very similar to the previous one, but a bit more cumbersome,
  because each element of a structure $\A$ has to play the role
  of several different elements in $μ(\A)$.
  Given any $Ξ\+(Ψ)$-sentence $ψ$ over $τ$,
  we need to construct a $Ξ\+(Φ)$-sentence $φ_ψ$ over $σ$,
  such that evaluating $φ_ψ$ on $\A$ is equivalent to evaluating $ψ$ on $μ(\A)$,
  for all $\A∈\C$.
  For the remainder of this proof,
  let $m$, $n$ be the parameters of the linear encoding $μ$.

  Again, we first deal with the kernel classes $Φ,Ψ$,
  and show the following claim by induction on the structure of $Ψ$-formulas:
  For every $Ψ$-sentence $ψ$ over \mbox{$τ_\at∪\Z$} and all \mbox{$h∈\range{n}$},
  with $\Z=\set{Z_1,…,Z_t}⊆\RelSyms[1]{\setminus}τ$,
  there is a $Φ$-sentence $φ_ψ^h$ over \mbox{$σ_\at∪\tilde{\Z}$},
  with $\tilde{\Z}=\set{Z_1^1,…,Z_t^n}⊆\RelSyms[1]{\setminus}σ$,\,
  such that
  \begin{align*}
    &\bigver{\A}{\at,\tuple{Z_r^j}_{r≤t}^{j≤n}}{a,\tuple{A_r^j}_{r≤t}^{j≤n}} \,⊨\, φ_ψ^h \\
    &\quad\text{\Iff} \\
    &\bigver{μ(\A)}{\at,\tuple{Z_r}_{r≤t}}{b,\tuple{B_r}_{r≤t}} \,⊨\, ψ, \\[1ex]
    &\quad\text{where $b$ is $\tuple{h,a}$ if $h≤m$, otherwise $h$,\, and} \\
    &\quad B_r =\; \smashoperator{\bigcup_{1≤j≤m}}\,\bigl(\set{j}{×}A_r^j\bigr)
    \,∪\, \,\smashoperator{\bigcup_{m<j≤n}}\, \setbuilder{j}{A_r^j=\domain{\A}},
  \end{align*}
  for all $\A∈\C$,\, $a∈\A$, and sets $\tuple{A_r^j}_{1≤r≤t}^{1≤j≤m}⊆\A$
  and $\tuple{A_r^j}_{1≤r≤t}^{m<j≤n}∈\set{∅,\domain{\A}}$.
  \begin{itemize}
  \item If $ψ = \at$, it suffices to set $φ_ψ^h = \at$.
  \item If $ψ = Z_r$, for some $Z_r∈\Z$,
    the translation is given by $φ_ψ^h = Z_r^h$.
  \item If $ψ = Q$, for some element symbol or set symbol $Q$ in $τ$,
    we use the fact that $μ$ allows for backward translation from $Ψ$ to $Φ$,
    and choose $φ_ψ^h$ to be the formula $φ_Q^h$,
    which is provided by \cref{def:backward-translation}~\ref{itm:bwd-set}.
    The definition asserts that
    this formula fulfills the induction hypothesis for the case
    where $\A$ and $μ(\A)$ are not extended using additional set symbols
    from $\tilde{\Z}$ and $\Z$.
    But since these symbols do not occur freely in $φ_Q^h$ and $Q$,
    their interpretations do not influence the evaluation of the formulas.
  \item If $ψ = ¬ψ_1$ \,or\, $ψ = ψ_1 ∨ ψ_2$,
    where $ψ_1$ and $ψ_2$ are formulas that satisfy the induction hypothesis,
    we set $φ_ψ^h = ¬φ_{ψ_1}^h$ \,or\, $φ_ψ^h = φ_{ψ_1}^h ∨ φ_{ψ_2}^h$,\, respectively.
  \item If $ψ = \dm[S]\tuple{ψ_i}_{i≤k}$,
    where $S$ is a relation symbol in $τ$ of arity $k+1≥2$,
    and $\tuple{ψ_i}_{i≤k}$ are $Ψ$-sentences over $τ_\at∪\Z$
    satisfying the hypothesis,
    we again rely on the premise
    that $μ$ allows for backward translation from $Ψ$ to $Φ$.
    We construct $φ_ψ^h$ by plugging the formulas $\tuple{φ_{ψ_i}^j}_{i≤k}^{j≤n}$
    provided by induction into the formula $φ_S^h$ of
    \cref{def:backward-translation}~\ref{itm:bwd-relation} as follows:
    \begin{equation*}
      φ_ψ^h = \bigver{φ_S^h}{\tuple{X_i^j}_{i≤k}^{j≤n}}{\tuple{φ_{ψ_i}^j}_{i≤k}^{j≤n}}.
    \end{equation*}
    For $1≤i≤k$ and $1≤j≤n$, let $A_i^{j\prime}$ be the set of elements $a'∈\A$
    that satisfy $φ_{ψ_i}^j$ in $\bigver{\A}{\tuple{Z_r^j}_{r≤t}^{j≤n}}{\tuple{A_r^j}_{r≤t}^{j≤n}}$,
    and let $B'_i$ be the set of elements $b'∈μ(\A)$
    that satisfy $ψ_i$ in $\bigver{μ(\A)}{\tuple{Z_r}_{r≤t}}{\tuple{B_r}_{r≤t}}$.
    The induction hypothesis ensures that
    \begin{equation*}
      B'_i =\; \smashoperator{\bigcup_{1≤j≤m}}\,\bigl(\set{j}{×}A_i^{j\prime}\bigr)
      \,∪\, \,\smashoperator{\bigcup_{m<j≤n}}\, \setbuilder{j}{A_i^{j\prime}=\domain{\A}}.
    \end{equation*}
    Hence, we obtain the required equivalence as follows:
    \begin{equation*}
      \begin{aligned}
        &\bigver{\A}{\at,\tuple{Z_r^j}_{r≤t}^{j≤n}}{a,\tuple{A_r^j}_{r≤t}^{j≤n}} \,⊨\, φ_ψ^h \\[0.5ex]
        \text{iff}\quad
        &\bigver{\A}{\at,\tuple{X_i^j}_{i≤k}^{j≤n}\:\!}{a,\tuple{A_i^{j\prime}}_{i≤k}^{j≤n}\!} \,⊨\, φ_S^h \\[0.5ex]
        \text{iff}\quad
        &\bigver{μ(\A)}{\at,\tuple{Y_i}_{i≤k}\:\!}{b,\tuple{B'_i}_{i≤k}} \,⊨\, \dm[S]\tuple{Y_i}_{i≤k} \\[0.5ex]
        \text{iff}\quad
        &\bigver{μ(\A)}{\at,\tuple{Z_r}_{r≤t}}{b,\tuple{B_r}_{r≤t}} \,⊨\, ψ.
      \end{aligned}
    \end{equation*}
  \item If $ψ = \bdm[S]\tuple{ψ_i}_{i≤k}$,
    supposing $Ψ$ includes backward modalities,
    we construct $φ_ψ^h$ using the same approach as in the previous case,
    the only difference being that we consider $\invS$ instead of $S$ and invoke
    \cref{def:backward-translation}~\ref{itm:bwd-backward} instead of
    \ref{def:backward-translation}~\ref{itm:bwd-relation}.
  \item If $ψ = \gdm ψ_1$, in case $Ψ$ includes global modalities,
    we again proceed as for the case $ψ = \dm[S]\tuple{ψ_i}_{i≤k}$,
    this time using~$\Tglob$ instead of $S$, with $k=1$,
    and referring to 
    \cref{def:backward-translation}~\ref{itm:bwd-global}.
  \end{itemize}

  Similarly to the proof of part~\ref{itm:forward-inclusion},
  we now extend the previous property to cover
  formulas with set quantifiers,
  evaluated on structures that may interpret the position symbol $\at$ arbitrarily.
  Our induction hypothesis is the following:
  For every $Ξ\+(Ψ)$-sentence $ψ$ over \mbox{$τ∪\Z$},
  with $\Z=\set{Z_1,…,Z_t}⊆\RelSyms[1]{\setminus}τ$ (possibly empty),
  there is a $Ξ\+(Φ)$-sentence $φ_ψ$ over \mbox{$σ∪\tilde{\Z}$},
  with $\tilde{\Z}=\set{Z_1^1,…,Z_t^n}⊆\RelSyms[1]{\setminus}σ$,\,
  such that
  \begin{align*}
    &\bigver{\A}{\tuple{Z_r^j}_{r≤t}^{j≤n}}{\tuple{A_r^j}_{r≤t}^{j≤n}} \,⊨\, φ_ψ \\
    &\quad\text{\Iff} \\
    &\bigver{μ(\A)}{\tuple{Z_r}_{r≤t}}{\tuple{B_r}_{r≤t}} \,⊨\, ψ, \quad\text{where} \\[1ex]
    &\quad B_r =\; \smashoperator{\bigcup_{1≤j≤m}}\,\bigl(\set{j}{×}A_r^j\bigr)
    \,∪\, \,\smashoperator{\bigcup_{m<j≤n}}\, \setbuilder{j}{A_r^j=\domain{\A}},
  \end{align*}
  for all structures $\A∈\C$, and sets $\tuple{A_r^j}_{1≤r≤t}^{1≤j≤m}⊆\A$
  and $\tuple{A_r^j}_{1≤r≤t}^{m<j≤n}∈\set{∅,\domain{\A}}$.
  \begin{itemize}
  \item If $ψ$ belongs to the kernel class $Ψ$,
    we apply the claim just proven,
    and construct $φ_ψ$ by substituting into the formula $φ_\ini$ provided by
    \cref{def:backward-translation}~\ref{itm:bwd-initial}:
    \mbox{$φ_ψ = \bigver{φ_\ini}{\tuple{X^j}^{j≤n}}{\tuple{φ_ψ^j}^{j≤n}}$}.
    Proceeding analogously to the proof of part~\ref{itm:forward-inclusion},
    we have to distinguish whether or not the position symbol $\at$ belongs to $τ$.
    (If it does not, $ψ$ is necessarily equivalent to $\gdm ψ$.)
    In both cases, the definition of $φ_\ini$ guarantees
    that the $\tilde{\Z}$-extended variant of $\A$ satisfies $φ_ψ$
    \Iff{} the $\Z$-extended variant of $μ(\A)$ satisfies $ψ$.
  \item If $ψ$ is a Boolean combination of subformulas
    that satisfy the induction hypothesis,
    then $φ_ψ$ is simply the corresponding Boolean combination 
    of the translated subformulas.
  \item If $ψ = \EE{Z_{t+1}}ψ_1$,
    where $ψ_1$ is a $Ξ\+(Ψ)$-sentence over \mbox{$τ∪\set{Z_1,…,Z_{t+1}}$}
    that satisfies the induction hypothesis,
    we choose $φ_ψ$ to be the formula
    \begin{equation*}
      \EE{\tuple{Z_{t+1}^j}^{j≤m}}
      \Bigl(\,
      \smashoperator{\bigvee_{N\,⊆\;\lrange[m\:\!\!]{\:\!\!n} \vphantom{\big(}}}
      \bigver{φ_{ψ_1}}{\bigtuple{Z_{t+1}^j}^{j>m}\!}{\!\bigtuple{N(j)}^{j>m}}
      \,\Bigr),
    \end{equation*}
    with $N(j)=⊤$ if $j∈N$, and $N(j)=⊥$ otherwise.
    For each set $N⊆\lrange[m]{n}$,
    let $φ_{ψ_1}^N$ denote the disjunct corresponding to $N$ in the formula above.
    By induction, we have the following equivalence:
    the interpretation
    $\map{\tuple{Z_{t+1}^j}^{j≤m}}{\tuple{A_{t+1}^j}^{j≤m}}$
    leads to satisfaction of $φ_{ψ_1}^N$ in the $\tilde{\Z}$-extended variant of $\A$
    \Iff{} 
    \begin{equation*}
      Z_{t+1} \,\mapsto\,\;
      \smashoperator{\bigcup_{1≤j≤m}}\,\bigl(\set{j}{×}A_{t+1}^j\bigr) \,∪\, \,N
    \end{equation*}
    is a satisfying choice for $ψ_1$ in the $\Z$-extended variant of $μ(\A)$.
    \qedhere
  \end{itemize}
\end{proof}

\subsection{Getting Rid of Multiple Edge Relations}
We now show how to encode a multi-relational graph into a $1$-relational one,
by inserting additional labeled elements
that represent the different edge relations.

\begin{proposition}
  \label{prp:multirelational-labeled}
  \NoMathBreak
  For all $t,u≥0$ and $Φ∈\set{\HG,\HBG}$,
  there is a linear encoding $μ$ from $\DIGRAPH[t,u]$ into $\DIGRAPH[t+u,1]$
  that allows for bidirectional translation within $Φ$.

  Moreover, $μ(\DIGRAPH[t,u])$ is $\PS{1}(\HG)$-definable over $\DIGRAPH[t+u,1]$.
\end{proposition}
\begin{proof}
  We choose $μ$ to be the linear encoding
  that assigns to each $t$-bit labeled, $u$-relational directed graph $\sD$
  the \mbox{$(t+u)$-bit} labeled ($1$-relational) directed graph $μ(\sD)=\sE$
  with domain $\range{u+1}×\domain{\sD}$,
  labeling sets $\inp{\sE}{P_i}=\set{1}×\inp{\sD}{P_i}$\!, for $1≤i≤t$,
  and $\inp{\sE}{P_{t+i}}=\set{i+1}×\domain{\sD}$, for $1≤i≤u$,
  and edge relation
  \begin{equation*}
    \begin{aligned}
      \inp{\sE}{R} = \; \smashoperator{\bigcup_{1≤i≤u}} \:\,
      \Bigl(&\bigsetbuilder{\bigtuple{\tuple{1,d},\tuple{i+1,d}}}{d∈\sD} \;\; ∪ \\[-2.6ex]
            &\bigsetbuilder{\bigtuple{\tuple{i+1,d},\tuple{1,d}}}{d∈\sD} \;\; ∪ \\[-0.8ex]
            &\bigsetbuilder{\bigtuple{\tuple{i+1,d},\tuple{i+1,d'}}}{\tuple{d,d'}∈\inp{\sD}{R_i}}\Bigr).
    \end{aligned}
  \end{equation*}
  That is,
  for each element $d∈\sD$ and for $1≤i≤u$,
  we introduce an additional element
  representing the “$R_i$-port” of $d$,
  and connect everything accordingly.

  Our forward translation,
  from $Φ$ on $\DIGRAPH[t,u]$ to $Φ$ on $μ(\DIGRAPH[t,u])$,
  is given by
  \begin{equation*}
    \begin{aligned}
      &\begin{aligned}
        ψ_{P_i}       &= P_i \quad\text{for $1≤i≤t$}, \\
        ψ_{R_i}       &= \dm\bigl(ψ_{i+1} ∧ \dm\dm(ψ_1 ∧ Y)\bigr) \quad\text{for $1≤i≤u$}, \\
        ψ_{\invr{R_i}} &= \dm\bigl(ψ_{i+1} ∧ \bdm\dm(ψ_1 ∧ Y)\bigr) \quad\text{for $1≤i≤u$}, \\
        \psiglob     &= \gdm(ψ_1 ∧ Y),\\
        ψ_\ini        &= \psiglob\+,
      \end{aligned} \\[1ex]
      &\begin{alignedat}{3}
        \text{where}&\; &    ψ_1 &= \textstyle ¬\bigvee_{1≤i≤u}(ψ_{i+1}) & \qquad&\text{(“regular”)}, \\
                    &   & ψ_{i+1} &= P_{t+i} \quad\text{for $1≤i≤u$}    &       &\text{(“$R_i$-port”)}.
      \end{alignedat}
    \end{aligned}
  \end{equation*}

  Our translation in the other direction,
  from $Φ$ on $μ(\DIGRAPH[t,u])$ to $Φ$ on $\DIGRAPH[t,u]$,
  is given by
  \begin{align*}
    φ_{P_i}^{h+1}   &= \begin{cases*}
                       P_i & for $h=0$ and $1≤i≤t$, \\
                       ⊥   & for $h=0$ and $t+1≤i≤t+u$, \\
                       ⊤   & for $1≤h≤u$ and $i=t+h$, \\
                       ⊥   & for $1≤h≤u$ and $i≠t+h$,
                     \end{cases*} \\
    φ_R^{h+1}      &= \begin{cases*}
                       \bigvee_{1≤i≤u}X^{i+1} & for $h=0$, \\
                       X^1 ∨ \dm[h]X^{h+1}   & for $1≤h≤u$,
                     \end{cases*} \\
    φ_\invR^{h+1}   &= \begin{cases*}
                       \bigvee_{1≤i≤u}X^{i+1} & for $h=0$, \\
                       X^1 ∨ \bdm[h]X^{h+1}  & for $1≤h≤u$,
                     \end{cases*} \\
    \phiglob[h+1] &= \gdm(X^1∨…∨X^{u+1}) \quad\;\,\text{for $0≤h≤u$}, \\
    φ_\ini         &= \phiglob[1].
  \end{align*}
  
  We can characterize $μ(\DIGRAPH[t,u])$ over the class $\DIGRAPH[t+u,1]$
  by the conjunction of the following $\PS{1}(\HG)$-definable properties,
  using our helper formulas $\tuple{ψ_i}_{1≤i≤u+1}$ 
  from the forward translation.
  \begin{itemize}
  \item A “port” that corresponds to a relation symbol $R_i$
    may not be associated with any other relation symbol $R_j$,
    nor be labeled with predicates $\tuple{P_j}_{1≤j≤t}$.
    \begin{equation*}
      \smashoperator{\bigwedge_{1≤i≤u}} \gbx\Bigl(
      ψ_{i+1} \,→\;
      ¬\smashoperator{\bigvee_{1≤j≤u,\;j≠i}}(ψ_{j+1}) \+∧\,
      ¬\smashoperator{\bigvee_{1≤j≤t}}(P_j) 
      \Bigr)
    \end{equation*}
  \item Every “regular element” is connected to its $u$ different “ports”,
    and nothing else.
    The uniqueness of each “$R_i$-port” can be expressed by
    the $\eqcl{\PS{1}(\HG)}$-formula $\seeone(ψ_{i+1})$,
    using the construction from \cref{ssec:singleton}.
    \begin{equation*}
      \gbx\Bigl(
      ψ_1 \,→\; 
      ¬\dm ψ_1 \+∧\+ 
      \smashoperator{\bigwedge_{1≤i≤u}}\seeone(ψ_{i+1})
      \Bigr)
    \end{equation*}
  \item Similarly,
    each “port” is connected to precisely one “regular element”
    and to an arbitrary number of “ports” of the same relation symbol,
    but not to any other ones.
    \begin{equation*}
      \smashoperator{\bigwedge_{1≤i≤u}} \gbx\Bigl(
      ψ_{i+1} \,→\; 
      \seeone(ψ_1) \+∧\,
      ¬\smashoperator{\bigvee_{1≤j≤u,\;j≠i}}\dm ψ_{j+1}
      \Bigr)
    \end{equation*}
  \item Finally,
    the links between “regular elements” and “ports”
    have to be bidirectional:
    for each edge from an element of one type
    to an element of a different type,
    the corresponding inverse edge must also exist.
    \begin{equation*}
      \smashoperator{\bigwedge_{1≤i≤u+1}}
      \AA{X} \gbx\bigl( ψ_i ∧\+ X \:→\: \bx(¬ψ_i → \dm X) \bigr)
    \end{equation*}
    Note that,
    in combination with the previous properties,
    this ensures that we have the same total number of elements
    for each type $i ∈ \range[1]{u+1}$.
    \qedhere
  \end{itemize}
\end{proof}

\subsection{Getting Rid of Vertex Labels}
Being able to eliminate multiple edge relations
at the cost of additional labeling sets
(see \cref{prp:multirelational-labeled}),
our natural next step is to encode labeled graphs into unlabeled ones.

\begin{proposition}
  \label{prp:labeled-unlabeled}
  For all $t≥1$ and $Φ∈\set{\HG,\HBG}$,
  there is a linear encoding $μ$ from $\DIGRAPH[t,1]$ into $\DIGRAPH$
  that allows for bidirectional translation within $Φ$.

  Moreover, $μ(\DIGRAPH[t,1])$ is $\PS{1}(\HG)$-definable over $\DIGRAPH$.
\end{proposition}
\begin{proof}
  We construct the linear encoding $μ$
  that assigns to each $t$-bit labeled directed graph $\sD$
  the (unlabeled) directed graph $μ(\sD)=\sE$
  with domain $(\set{1}×\domain{\sD}) ∪ \range[2]{t+3}$
  and edge relation
  \begin{align*}
    \inp{\sE}{R} =
    &\phantom{{}∪\:{}}\bigsetbuilder{\bigtuple{\tuple{1,d},\tuple{1,d'}}}{\tuple{d,d'}∈\inp{\sD}{R}} \\
    &∪\:\bigsetbuilder{\bigtuple{\tuple{1,d},\,3}}{d∈\sD} \\
    &∪\:\:\smashoperator{\bigcup_{1≤i≤t}}\; \bigsetbuilder{\bigtuple{\tuple{1,d},\,i+3}}{d∈\inp{\sD}{P_i}} \\[-0.2ex]
    &∪\:\bigsetbuilder{\tuple{i+3,\:i+2}}{1≤i≤t} \\
    &∪\:\bigsetbuilder{\tuple{i+3,\:2}}{0≤i≤t}.
  \end{align*}
  The idea is to introduce a gadget that contains a separate element
  for each labeling set of the original graph,
  and then connect the “regular elements” to this gadget
  in a way that corresponds to their respective labeling.
  The gadget is easily identifiable because
  it contains the only element in the graph that has no outgoing edge
  (namely, element~$2$).
  We ensure this by connecting all the “regular elements” to element~$3$.

  Our forward translation,
  from $Φ$ on $\DIGRAPH[t,1]$ to $Φ$ on $μ(\DIGRAPH[t,1])$,
  is given by
  \begin{gather*}
    \begin{aligned}
      ψ_{P_i}   &= \dm ψ_{i+3} \quad\text{for $1≤i≤t$}, \\
      ψ_R      &= \dm(ψ_1 ∧ Y), \\
      ψ_\invR   &= \bdm Y, \\
      \psiglob &= \gdm(ψ_1 ∧ Y),\\
      ψ_\ini    &= \psiglob\+,
    \end{aligned} \\[1ex]
    \begin{alignedat}{2}
      \text{where\!}& &    ψ_1 &= \textstyle ¬\bigvee_{2≤i≤t+3}(ψ_i), \\
                    & &    ψ_2 &= \bx ⊥, \\
                    & &    ψ_3 &= \dm ψ_2 ∧ \bx ψ_2, \\
                    & & ψ_{i+3} &= \dm ψ_2 ∧ \dm ψ_{i+2} ∧ \bx(ψ_2 ∨ ψ_{i+2}) \\
                    & &        &\quad\:\, \text{for $1≤i≤t$}.
    \end{alignedat}
  \end{gather*}

  Our translation in the other direction,
  from $Φ$ on $μ(\DIGRAPH[t,1])$ to $Φ$ on $\DIGRAPH[t,1]$,
  is given by
  \begin{equation*}
    \begin{aligned}
      φ_R^h       &= \begin{cases*}
                       \dm X^1 \,∨\, X^3 \;∨{} \\[-1ex]
                       \quad\;\bigvee_{1≤i≤t}(P_i∧X^{i+3}) & for $h=1$, \\
                       ⊥            & for $h=2$, \\
                       X^2          & for $h=3$, \\
                       X^2 ∨ X^{h-1} & for $4≤h≤t+3$,
                     \end{cases*} \\
      φ_\invR^h    &= \begin{cases*}
                       \bdm X^1                    & for $h=1$, \\
                       \bigvee_{0≤i≤t}X^{i+3}        & for $h=2$, \\
                       \gdm X^1 ∨ X^{h+1}           & for $h=3$, \\
                       \gdm(P_{h-3} ∧ X^1) ∨ X^{h+1} \hspace{-1ex}& for $4≤h≤t+2$, \\
                       \gdm(P_{h-3} ∧ X^1)          & for $h=t+3$,
                     \end{cases*} \\
      \phiglob[h] &= \gdm(X^1∨…∨X^{t+3}) \quad\+\text{for $1≤h≤t+3$}, \\
      φ_\ini       &= \phiglob[1].
    \end{aligned}
  \end{equation*}

  Using the helper formulas $\tuple{ψ_i}_{1≤i≤t+3}$ from the forward translation,
  we can characterize $μ(\DIGRAPH[t,1])$ over $\DIGRAPH$ as
  \begin{equation*}
    \gdm ψ_1
    \,∧\, \smashoperator{\bigwedge_{2≤i≤t+3}}\totone(ψ_i)
    \,∧\, \gbx(ψ_1→ \dm ψ_3 ∧ ¬\dm ψ_2).
  \end{equation*}
  Here, each $\eqcl{\PS{1}(\HG)}$-subformula \,$\totone(ψ_i)$
  is obtained through
  the singleton construction from \cref{ssec:singleton}.
\end{proof}

\subsection{Getting Rid of Backward Modalities}
For the sake of completeness,
we also describe the encoding
that lets us simulate backward modalities
by means of an additional edge relation.

\begin{proposition}
  \label{prp:hbg-to-hg}
  There is a linear encoding $μ$ from $\DIGRAPH$ into $\DIGRAPH[0,2]$
  that allows for bidirectional translation \mbox{between} $\HBG$ and $\HG$.

  Moreover, $μ(\DIGRAPH)$ is $\PS{1}(\HG)$-definable over $\DIGRAPH[0,2]$.
\end{proposition}
\begin{proof}
  The encoding is straightforward:
  to each directed graph $\sD$,
  we assign a copy $μ(\sD) = \sE$
  that is enriched with a second edge relation,
  which coincides with the inverse of the first.
  Formally,
  $\domain{\sE} = \set{1} × \domain{\sD}$,
  \begin{align*}
    \inp{\sE}{R_1} &= \bigsetbuilder{\bigtuple{\tuple{1,d},\tuple{1,d'}}}{\tuple{d,d'}∈\inp{\sD}{R}},
    \quad \text{and} \\
    \inp{\sE}{R_2} &= \bigsetbuilder{\tuple{e'\!,e}}{\tuple{e,e'}∈\inp{\sE}{R_1}}.
  \end{align*}

  With this,
  in order to translate between $\HBG$ on $\DIGRAPH$ and $\HG$ on $μ(\DIGRAPH)$,
  we merely have to replace backward modalities by $R_2$-modalities,
  and vice versa.
  Hence, when we fix our forward translation,
  we choose $ψ_R = \dm[1] Y$ and $ψ_\invR = \dm[2] Y$,
  and for the backward translation
  we set $φ_{R_1}^1 = \dm X^1$ and $φ_{R_2}^1 = \bdm X^1$\!.

  To define $μ(\DIGRAPH)$ over $\DIGRAPH[0,2]$,
  we can use the following $\PS{1}(\HG)$-formula:
  \begin{equation*}
    \AA{X}\gbx\bigl(X → \bx[1]\dm[2]X ∧ \bx[2]\dm[1]X\bigr)
    \qedhere
  \end{equation*}
\end{proof}

\subsection{Getting Rid of Directed Edges}
In order to encode a directed graph into an undirected one,
we proceed in a similar manner to the elimination of multiple edge relations
in \cref{prp:multirelational-labeled}.
Using an ad-hoc trick,
we can do this by introducing only one additional labeling set.

\begin{proposition}
  \label{prp:digraph-1bitgraph}
  There is a linear encoding $μ$ from $\DIGRAPH$ into $\GRAPH[1,1]$
  that allows for bidirectional translation \mbox{between} $\HBG$ and $\HG$.

  Moreover, $μ(\DIGRAPH)$ is $\PS{1}(\HG)$-definable over $\GRAPH[1,1]$.
\end{proposition}
\begin{proof}
  A suitable choice for $μ$ is to take the function that
  assigns to every directed graph $\sD$
  the \mbox{$1$-bit} labeled undirected graph $μ(\sD)=\sG$
  with domain $(\range{3}×\domain{\sD}) ∪ \range[4]{6}$,\,
  labeling set $\inp{\sG}{P} = \range[4]{6}$,
  and edge relation $\inp{\sG}{R}=\bigsetbuilder{\tuple{g,g'}}{\set{g,g'}∈G}$,\,
  where
  \begin{align*}
    G =
    &\phantom{{}∪\:{}}\bigsetbuilder{\set{\tuple{1,d},\tuple{2,d}}}{d∈\sD} \\
    &∪\:\bigsetbuilder{\set{\tuple{1,d},\tuple{3,d}}}{d∈\sD} \\
    &∪\:\bigsetbuilder{\set{\tuple{2,d},\tuple{3,d'}}}{\tuple{d,d'}∈\inp{\sD}{R}} \\
    &∪\:\bigsetbuilder{\set{\tuple{2,d},\,4}}{d∈\sD} \\
    &∪\:\bigsetbuilder{\set{\tuple{3,d},\,5}}{d∈\sD} \\
    &∪\:\bigset{\set{5,\,6}}.
  \end{align*}
  The idea is that we connect each original element ${d∈\sD}$
  to two new elements,
  which represent the “outgoing port” and “incoming port” of $d$,
  and use undirected edges between “ports”
  to simulate directed edges between “regular elements”.
  In order to distinguish the different types of elements,
  we connect them in different ways
  to the additional $P$-labeled elements.

  Our forward translation,
  from $\HBG$ on $\DIGRAPH$ to $\HG$ on $μ(\DIGRAPH)$,
  is given by
  \begin{gather*}
    \begin{aligned}
      ψ_R      &= \dm\bigl(ψ_2 ∧ \dm\dm(ψ_1 ∧ Y)\bigr), \\
      ψ_\invR   &= \dm\bigl(ψ_3 ∧ \dm\dm(ψ_1 ∧ Y)\bigr), \\
      \psiglob &= \gdm(ψ_1 ∧ Y),\\
      ψ_\ini    &= \psiglob\+,
    \end{aligned} \\[1ex]
    \begin{alignedat}{2}
      \text{where} \quad ψ_1 &= ¬(ψ_2 ∨ … ∨ ψ_6) & &\text{(“regular”)}, \\
                         ψ_2 &= \dm ψ_4          & &\text{(“outgoing”)}, \\
                         ψ_3 &= ¬P ∧ \dm ψ_5     & &\text{(“incoming”)}, \\
                         ψ_4 &= P ∧ ¬\dm P ∧ \phantom{¬}\dm ¬P, \!\! \\
                         ψ_5 &= P ∧ \phantom{¬}\dm P ∧ \phantom{¬}\dm ¬P, \!\! \\
                         ψ_6 &= P ∧ \phantom{¬}\dm P ∧ ¬\dm ¬P. \!\!
    \end{alignedat}
  \end{gather*}

  Our backward translation,
  from $\HG$ on $μ(\DIGRAPH)$ to $\HBG$ on $\DIGRAPH$,
  is given by
  \begin{align*}
    φ_P^h       &= \begin{cases*}
                       ⊥   & for $1≤h≤3$, \\
                       ⊤   & for $4≤h≤6$,
                   \end{cases*} \\
    φ_R^h       &= \begin{cases*}
                     X^2 ∨ X^3             & for $h=1$, \\
                     X^1 ∨ \dm X^3 ∨ X^4   & for $h=2$, \\
                     X^1 ∨ \bdm X^2 ∨ X^5  & for $h=3$, \\
                     \gdm X^2              & for $h=4$, \\
                     \gdm X^3 ∨ X^6        & for $h=5$, \\
                     X^5                   & for $h=6$,
                   \end{cases*} \\
    \phiglob[h] &= \gdm(X^1∨…∨X^6) \quad\text{for $1≤h≤6$}, \\
    φ_\ini       &= \phiglob[1].
  \end{align*}

  We can define $μ(\DIGRAPH)$ over $\GRAPH[1,1]$
  with the following $\eqcl{\PS{1}(\HG)}$-formula.
  It makes use of
  our helper formulas $\tuple{ψ_i}_{1≤i≤6}$ from the forward translation
  and the constructions \,$\seeone(ψ_i)$\, and \,$\totone(ψ_i)$\,
  from \cref{ssec:singleton}.
  \begin{equation*}
    \begin{aligned}
      &\smashoperator{\bigwedge_{4≤i≤6}} \totone(ψ_i) \,\,∧ {} \\
      &\gbx\bigl( ψ_2 \,→\, \seeone(ψ_1) ∧ \bx(ψ_1 ∨ ψ_3 ∨ ψ_4) \bigr) \,\,∧ {} \\
      &\gbx\bigl( ψ_3 \,→\, \seeone(ψ_1) ∧ \bx(ψ_1 ∨ ψ_2 ∨ ψ_5) \bigr) \,\,∧ {} \\
      &\gbx\bigl( ψ_1 \,→\, \seeone(ψ_2) ∧ \seeone(ψ_3) ∧ \bx(ψ_2 ∨ ψ_3) \bigr)
    \end{aligned}
  \end{equation*}
  The first line states that the three $P$-labeled elements are unique,
  which forces $5$ and $6$ to be connected.
  The remaining lines ensure
  that each “port” is connected to exactly one “regular element”,
  and, conversely,
  that every “regular element” is linked to precisely
  one “outgoing port” and one “incoming port”.
  As a consequence,
  the number of “regular elements” must be the same
  as the number of “ports” of each type.
  Furthermore,
  the formula restricts the types of neighbors each element can have,
  while the usage of the helper formulas $ψ_2$ and $ψ_3$
  makes sure that the required connections to the $P$-labeled elements
  are established.
  Finally,
  the fact that $ψ_1$ characterizes
  the “regular elements” as the “remaining ones”
  guarantees that there are no unaccounted-for elements.
\end{proof}

\subsection{Getting Rid of Global Modalities}
Our last encoding function lets us simulate global modalities
by inserting a new element
that is bidirectionally connected to all the “regular elements”.

\begin{proposition}
  \label{prp:digraph-pdigraph}
  There is a linear encoding $μ$ from $\DIGRAPH$ into $\PDIGRAPH$
  that allows for bidirectional translation \mbox{between} $\HG$ and $\H$.

  Furthermore, $μ$ can be easily adapted into
  a linear encoding $μ'$ from $\DIGRAPH$ into $\DIGRAPH$
  that satisfies the following figurative inclusions,
  for arbitrary $ℓ≥2$:
  \begin{alignat*}{2}
    \sem[\DIGRAPH]{\SS{ℓ}(\HG)} &\,\figsubeq{μ'}\;\, &&\sem[\DIGRAPH]{\gbx\SS{ℓ}(\H)} \,, \\
    \sem[\DIGRAPH]{\PS{ℓ}(\HG)} &\,\figeq{μ'}\;\,    &&\sem[\DIGRAPH]{\gbx\PS{ℓ}(\H)} \,.
    \;\;\;\qquad\qedhere
  \end{alignat*}
\end{proposition}
\begin{proof}
  We choose $μ$ to be the linear encoding
  that maps each directed graph $\sD$
  to the pointed directed graph $μ(\sD)=\sE$
  with domain $(\set{1}×\domain{\sD}) ∪ \range[2]{3}$,\,
  position $\inp{\sE}{\at} = 2$,
  and edge relation
  \begin{align*}
    \inp{\sE}{R} =
    &\phantom{{}∪\:{}}\bigsetbuilder{\bigtuple{\tuple{1,d},\tuple{1,d'}}}{\tuple{d,d'}∈\inp{\sD}{R}} \\
    &∪\:\bigsetbuilder{\bigtuple{\tuple{1,d},\,2}}{d∈\sD} \\
    &∪\:\bigsetbuilder{\bigtuple{2,\,\tuple{1,d}}}{d∈\sD} \\
    &∪\:\bigset{\tuple{2,\,3}}.
  \end{align*}
  One can distinguish element~$2$ from the others
  because it is connected to~$3$,
  which is the only element without any outgoing edge.

  Our forward translation,
  from $\HG$ on $\DIGRAPH$ to $\H$ on $μ(\DIGRAPH)$,
  is given by
  \begin{gather*}
    \begin{aligned}
      ψ_R      &= \dm(ψ_1 ∧ Y), \\
      \psiglob &= \dm(ψ_2 ∧\+ \dm(ψ_1 ∧ Y)),\\
      ψ_\ini    &= \dm(ψ_1 ∧ Y),
    \end{aligned} \\[1ex]
    \begin{gathered}
      \text{where}\quad ψ_1 = \dm\dm\bx ⊥ \quad\text{and}\quad ψ_2 = \dm\bx ⊥.
    \end{gathered}
  \end{gather*}

  Our backward translation,
  from $\H$ on $μ(\DIGRAPH)$ to $\HG$ on $\DIGRAPH$,
  is given by
  \begin{align*}
    φ_R^h &= \begin{cases*}
               \dm X^1 ∨ X^2  & for $h=1$, \\
               \gdm X^1 ∨ X^3 & for $h=2$, \\
               ⊥              & for $h=3$,
             \end{cases*} \\
    φ_\ini &= \gdm X^2.
  \end{align*}

  Turning to the second claim of the \lcnamecref{prp:digraph-pdigraph},
  we obtain $μ'(\sD)$ by simply removing the position marker from $μ(\sD)$,
  i.e., for every directed graph $\sD$,\,
  $μ'(\sD)$ is such that $\ver{μ'(\sD)}{\at}{2} = μ(\sD)$.

  For the forward figurative inclusions,
  let $Ξ ∈ \set{\SS{ℓ},\, \PS{ℓ}}$,\, for some arbitrary $ℓ≥0$.
  By applying \cref{lem:translation-inclusion}~\ref{itm:forward-inclusion}
  on $μ$, we get that
  for every $Ξ\+(\HG)$-sentence $φ$ over $\set{R}$,
  there is a $Ξ\+(\H)$-sentence $ψ_φ$ over $\set{\at,R}$
  such that, for all $\sD∈\DIGRAPH$,
  \begin{alignat*}{1}
    \sD \,⊨\, φ \quad&\text{iff}\quad \ver{μ'(\sD)}{\at}{2} \,⊨\, ψ_φ\+, \\
                     &\text{iff}\quad μ'(\sD) \,⊨\, \gbx(ψ_2 → ψ_φ).
  \end{alignat*}
  Hence,\,
  $\sem[\DIGRAPH]{\+Ξ\+(\HG)} \figsubeq{μ'} \sem[\DIGRAPH]{\gbx\,Ξ\+(\H)}\+$.

  For the backward figurative inclusion,
  we require that $ℓ≥2$.
  Slightly adapting the proof of
  \cref{lem:translation-inclusion}~\ref{itm:backward-inclusion}
  to discard the part where we make use of the formula $φ_\ini$ from
  \cref{def:backward-translation}~\ref{itm:bwd-initial}
  (incidentally allowing us to merge the two consecutive induction proofs),
  it is easy to show the following:
  Given $h∈\range{3}$ and any $\PS{ℓ}(\H)$-sentence $ψ$ over $\set{\at,R}$,
  we can construct a $\PS{ℓ}(\HG)$-sentence $φ_ψ^h$ over $\set{\at,R}$
  such that, for all $\sD∈\DIGRAPH$ and $d∈\sD$,
  \begin{align*}
    &\ver{\sD}{\at}{d} \,⊨\, φ_ψ^h \quad\text{iff}\quad \ver{μ'(\sD)}{\at}{e} \,⊨\, ψ, \\[0.5ex]
    &\quad\text{where $e$ is $\tuple{h,d}$ if $h=1$, and $h$ otherwise.}
  \end{align*}
  This immediately gives us a way of translating $\gbx ψ$:
  \begin{equation*}
    \sD \,⊨\, \gbx(φ_ψ^1 ∧ φ_ψ^2 ∧ φ_ψ^3) \quad\text{iff}\quad μ'(\sD) \,⊨\, \gbx ψ.
  \end{equation*}
  The left-hand side sentence can be transformed into prenex normal form
  by simulating the global box with a universal set quantifier.
  Checking that a given set is \emph{not} a singleton can be done in $\SS{1}(\HG)$,
  since the negation is $\PS{1}(\HG)$-expressible
  (see \cref{ssec:singleton}).
  Thus,
  the given formula is equivalent to a $\PS{ℓ}(\HG)$-formula,
  and we obtain that\,
  $\vphantom{\Bigl(}
  \sem[\DIGRAPH]{\PS{ℓ}(\HG)} \figsupeq{μ'} \sem[\DIGRAPH]{\gbx\PS{ℓ}(\H)}\+$.
\end{proof}

%% file: tex/acknowledgments.tex
\section*{Acknowledgments}
\addcontentsline{toc}{section}{Acknowledgments}

I would like to thank Olivier Carton,
my PhD supervisor,
for all the pleasant and helpful discussions
that lead to the present paper.
Furthermore,
I am grateful to Charles Paperman and
an anonymous reviewer of my previous paper,
who both suggested that there might be a connection
between distributed graph automata and modal logic.
Their intuition was spot-on!
\;{\Large\smiley{}}

%% file: ml-paper.bbl
\begin{thebibliography}{9}
\bibitem[AC06]{AC06}
  C.~Areces, B.~ten~Cate (2006),
  \emph{Hybrid Logics.}
  In P.~Blackburn, J.~van~Benthem, F.~Wolter, editors,
  Handbook of Modal Logic: 821--868
\bibitem[Ben83]{Ben83}
  J.~van~Benthem (1983),
  \emph{Modal Logic and Classical Logic.}
  Bibliopolis
\bibitem[Bul69]{Bul69}
  R.~A.~Bull (1969),
  \emph{On Modal Logic with Propositional Quantifiers.}
  J.~Symb.~Log.~34(2): 257--263
\bibitem[BRV02]{BRV02}
  P.~Blackburn, M.~de~Rijke, Y.~Venema (2002),
  \emph{Modal Logic.}
  Cambridge University Press
\bibitem[Cat06]{Cat06}
  B.~ten~Cate (2006),
  \emph{Expressivity of Second Order Propositional Modal Logic.}
  J.~Philosophical Logic 35(2): 209--223
\bibitem[Fin70]{Fin70}
  K.~Fine (1970),
  \emph{Propositional quantifiers in modal logic.}
  Theoria 36(3): 336--346
\bibitem[GR92]{GR92}
  D.~Giammarresi, A.~Restivo (1992),
  \emph{Recognizable Picture Languages.}
  IJPRAI 6(2\&3): 241--256
\bibitem[GRST96]{GRST96}
  D.~Giammarresi, A.~Restivo, S.~Seibert, W.~Thomas (1996),
  \emph{Monadic Second-Order Logic over Rectangular Pictures and Recognizability by Tiling Systems.}
  Inf.~Comput.~125(1): 32--45
\bibitem[H+12]{H+12}
  L.~Hella, M.~Järvisalo, A.~Kuusisto, J.~Laurinharju, T.~Lempiäinen, K.~Luosto, J.~Suomela, J.~Virtema (2012),
  \emph{Weak models of distributed computing, with connections to modal logic.}
  PODC 2012: 185--194;
  \href{http://arxiv.org/abs/1205.2051}{ arXiv:1205.2051 }
\bibitem[H+15]{H+15}
  L.~Hella, M.~Järvisalo, A.~Kuusisto, J.~Laurinharju, T.~Lempiäinen, K.~Luosto, J.~Suomela, J.~Virtema (2015),
  \emph{Weak models of distributed computing, with connections to modal logic.}
  Distributed Computing 28(1): 31--53
\bibitem[Kuu08]{Kuu08}
  A.~Kuusisto (2008),
  \emph{A modal perspective on monadic second-order alternation hierarchies.}
  Advances in Modal Logic 2008: 231--247
\bibitem[Kuu15]{Kuu15}
  A.~Kuusisto (2015),
  \emph{Second-order propositional modal logic and monadic alternation hierarchies.}
  Ann.~Pure~Appl.~Logic 166(1): 1--28
\bibitem[Mat02]{Mat02}
  O.~Matz (2002),
  \emph{Dot-depth, monadic quantifier alternation, and first-order closure over grids and pictures.}
  Theor.~Comput.~Sci.~270(1-2): 1--70
\bibitem[MST02]{MST02}
  O.~Matz, N.~Schweikardt, W.~Thomas (2002),
  \emph{The Monadic Quantifier Alternation Hierarchy over Grids and Graphs.}
  Inf.~Comput.~179(2): 356--383
\bibitem[MT97]{MT97}
  O.~Matz, W.~Thomas (1997),
  \emph{The Monadic Quantifier Alternation Hierarchy over Graphs is Infinite.}
  LICS~1997: 236--244
\bibitem[Rei15]{Rei15}
  F.~Reiter (2015),
  \emph{Distributed Graph Automata.}
  LICS~2015: 192-201;
  see also \href{http://arxiv.org/abs/1408.3030}{ arXiv:1408.3030 }
\bibitem[Sch97]{Sch97}
  N.~Schweikardt (1997),
  \emph{The Monadic Quantifier Alternation Hierarchy over Grids and Pictures.}
  CSL 1997: 441--460
\end{thebibliography}
